\documentclass[12pt,a4paper,oneside]{article}
\usepackage{graphicx,amssymb,amsmath,color}
\usepackage[colorlinks=true,urlcolor=blue,anchorcolor=blue,citecolor=red,filecolor=blue
,linkcolor=blue,menucolor=blue,linktocpage=true,pdfproducer=medialab,pdfa=true
]{hyperref}
\usepackage{cite}
\usepackage{relsize}
\usepackage{enumerate}
\usepackage[margin=2.3 cm]{geometry}
\usepackage{slashed}
\usepackage{multirow}
\usepackage[normalem]{ulem}
\usepackage[dvipsnames, x11names]{xcolor}
\usepackage[square,comma,sort&compress,numbers]{natbib}
\usepackage{soul}

\usepackage{cleveref}
\usepackage{cancel}

\def\beq{\begin{equation}}
\def\eeq{\end{equation}}
\def\bea{\begin{eqnarray}}
\def\eea{\end{eqnarray}}

\def\bit{\begin{itemize}}
\def\eit{\end{itemize}}

\def\baa{\begin{array}}
\def\eaa{\end{array}}

\def\d{\partial}

\def\simgt{\mathrel{\lower2.5pt\vbox{\lineskip=0pt\baselineskip=0pt
           \hbox{$>$}\hbox{$\sim$}}}}
\def\simlt{\mathrel{\lower2.5pt\vbox{\lineskip=0pt\baselineskip=0pt
           \hbox{$<$}\hbox{$\sim$}}}}

\def\d{\mathrm{d}}

\def\O{\mathcal{O}}
\def\vec{\mathbf}

\def\P{\mathcal{P}}

\def\vecp{\vec{p}}

\def\bfc{\begin{figure}\begin{center}}
\def\efc{\end{center}\end{figure}}

\newcommand{\overbar}[1]{\mkern 1.5mu\overline{\mkern-1.5mu#1\mkern-1.5mu}\mkern 1.5mu}

\newcommand{\blue}[1]{{\color{blue} #1}}

\definecolor{chromeyellow}{rgb}{1.0, 0.65, 0.0}
\definecolor{darkcoral}{rgb}{0.8, 0.36, 0.27}
\definecolor{cadmiumgreen}{rgb}{0.0, 0.42, 0.24}

\begin{document}

\begin{flushright}
\hspace{3cm} 
{
\small
KCL-PH-TH/2024-01
}
\end{flushright}
\vspace{.6cm}
\begin{center}

\hspace{-0.4cm}{\Large \bf 
Criterion for ultra-fast bubble walls: the impact of hydrodynamic obstruction
}

\vspace{1cm}{}
\end{center}

\begin{center}
Wen-Yuan Ai,$^{*1}$ Xander Nagels$^{\dagger 2}$ and Miguel Vanvlasselaer$^{\ddagger 2}$ \\
\vskip0.4cm

{\it $^1$Theoretical Particle Physics and Cosmology, King’s College London,\\ Strand, London WC2R 2LS, UK}\par
{\it $^2$ Theoretische Natuurkunde and IIHE/ELEM, Vrije Universiteit Brussel,
\& The International Solvay Institutes, Pleinlaan 2, B-1050 Brussels, Belgium }
\vskip1.cm
\end{center}

\bigskip \bigskip \bigskip

%%%%%%%%%%%%%%%%%%%%%%%%%%%%%%%%%%%%%%%%%%%%%%%%%%%%%%%%%%%%%%%%%%%%%%%%%%
\begin{abstract}

The B\"{o}deker-Moore thermal friction~\cite{Bodeker:2009qy} is usually used to determine whether or not a bubble wall can run away. However, the friction on the wall is not necessarily a monotonous function of the wall velocity and could have a maximum before it reaches the B\"{o}deker-Moore limit. In this paper, we compare the maximal hydrodynamic obstruction, a frictional force that exists in local thermal equilibrium, and the B\"{o}deker-Moore thermal friction.  We study the former in a fully analytical way, clarifying its physical origin and providing a simple expression for its corresponding critical phase transition strength above which the driving force cannot be balanced out by the maximal hydrodynamic obstruction. We find that for large parameter space, the maximal hydrodynamic obstruction is larger than the B\"{o}deker-Moore thermal friction, indicating that the conventional criterion for the runaway behavior of the bubble wall may have to be modified. We also explain how to apply efficiently the modified criterion to particle physics models and discuss possible limitations of the analysis carried out in this paper. 
\end{abstract}

\vfill
\noindent\line(1,0){188}
{\footnotesize{ \\ %E-mail:
\text{$^*$~wenyuan.ai@kcl.ac.uk}\\
\text{$^\dagger$~xander.staf.a.nagels@vub.be}   }\\
\text{$^\ddagger$~miguel.vanvlasselaer@vub.be}}

\newpage

\hrule
\tableofcontents
\vskip.8cm
\hrule

\section{Introduction}

Cosmological first-order phase transitions (FOPTs) have far-reaching phenomenological consequences. The collisions and mergers of the nucleated bubbles could result in a stochastic background of gravitational waves (GWs)~\cite{Witten:1984rs,Kosowsky:1991ua,Kosowsky:1992vn,Kamionkowski:1993fg,Hindmarsh:2013xza,Caprini:2015zlo,Caprini:2019egz}.\footnote{Excitingly, such an SGWB has been reported by several Pulsar Timing Array projects~\cite{NANOGrav:2023gor,EPTA:2023fyk,Reardon:2023gzh,Xu:2023wog} recently, whose source however may prefer a supermassive black hole explanation.}   The bubbles may also be pivotal for generating the cosmic matter-antimatter asymmetry~\cite{Kuzmin:1985mm,Shaposhnikov:1987tw,Morrissey:2012db,Garbrecht:2018mrp}.  Renewed interest in FOPTs in the early Universe has been triggered by the recent approval of the Laser Interferometer Space Antenna (LISA) project~\cite{LISA:2017pwj} and the detection of GWs~\cite{LIGOScientific:2016aoc}. Besides LISA, many other space-based GW detectors have been proposed, such as Big Bang Observer (BBO)~\cite{Corbin:2005ny}, Deci-hertz Interferometer Gravitational wave Observatory (DECIGO)~\cite{Kawamura:2011zz}, Taiji~\cite{Gong:2014mca}, and TianQin~\cite{TianQin:2015yph}.

FOPTs may also be relevant for producing primordial black holes~\cite{Kodama:1982sf,Hawking:1982ga,Garriga:2015fdk,Deng:2017uwc,Gross:2021qgx,Baker:2021nyl,Kawana:2021tde,Liu:2021svg,Jung:2021mku,Huang:2022him,Lewicki:2023ioy,Gouttenoire:2023naa}, particle dark matter~\cite{Baker:2019ndr,Chway:2019kft,Chao:2020adk}, and intergalactic magnetic fields~\cite{Vachaspati:1991nm,Olea-Romacho:2023rhh}. All the above-mentioned processes depend crucially on the regime of the bubble wall expansion, making the terminal velocity $\xi_w$ a key parameter. In recent years, ultrarelativistic bubble walls (bubbles expanding with a Lorentz factor $\gamma_w \equiv 1/\sqrt{1-\xi_w^2} \gg 1$) have been used for phenomenological studies on baryogenesis~\cite{Cline:2020jre,Azatov:2021irb,Baldes:2021vyz,Huang:2022vkf,Chun:2023ezg,Dichtl:2023xqd} and dark matter production~\cite{Azatov:2021ifm,Azatov:2022tii,Borah:2022cdx,Baldes:2023fsp,Giudice:2024tcp}. For the application of ultrarelativistic bubble walls, it is extremely important to determine whether the bubble wall can enter into the ultrarelativistic regime or not~\cite{Banerjee:2024qiu}. Furthermore, runaway and non-runaway walls could have very different GW signals (see Refs.~\cite{Caprini:2015zlo, Caprini:2018mtu,Caprini:2019egz, LISACosmologyWorkingGroup:2022jok,Athron:2023xlk,Roshan:2024qnv} for reviews and see, e.g., Refs.~\cite{Jinno:2019jhi, Lewicki:2022pdb} for studies on supercooled PTs). 

Estimates of the terminal wall velocity are usually based on kinetic theory~\cite{Dine:1992wr,Liu:1992tn,Moore:1995ua,Moore:1995si}, improvement of it~\cite{Cline:2020jre, Cline:2021iff,Laurent:2022jrs}, or holographic methods for strongly coupled theories~\cite{Bea:2021zsu, Baldes:2020kam,Bigazzi:2020avc,Bigazzi:2021ucw,Bea:2022mfb,Li:2023xto,Wang:2023lam} (see Ref.~\cite{Kang:2024xqk} for a quasiparticle (quasigluons) method). Typically, calculating the terminal wall velocity is a very challenging task, and has only been performed for a  limited number of models~\cite{Liu:1992tn,Moore:1995si,Moore:1995ua,Dorsch:2018pat,Laurent:2022jrs,Jiang:2022btc}. In such computations, one needs to solve the scalar field equation of motion (EoM) coupled to the Boltzmann equations describing the particles in the plasma. Simplifications can be obtained by assuming local thermal equilibrium (LTE)~\cite{Konstandin:2010dm, BarrosoMancha:2020fay,Balaji:2020yrx, Ai:2021kak,Wang:2022txy,Ai:2023see,Krajewski:2023clt, Sanchez-Garitaonandia:2023zqz} or in the ultrarelativistic regime~\cite{Bodeker:2009qy, Bodeker:2017cim,Hoche:2020ysm, Azatov:2020ufh,Gouttenoire:2021kjv, Ai:2023suz,Azatov:2023xem}.  In LTE, the wall velocity can be determined with the help of an additional matching condition for the hydrodynamic quantities on both sides of the wall~\cite{Ai:2021kak,Ai:2023see}. The LTE approximation is shown to work well for strongly coupled theories~\cite{Sanchez-Garitaonandia:2023zqz} whose FOPT-related phenomenology has been studied in, e.g., Refs.~\cite{Schwaller:2015tja,Aoki:2017aws,Helmboldt:2019pan,Agashe:2019lhy,Bigazzi:2020avc,Halverson:2020xpg,Huang:2020crf,Ares:2020lbt,Kang:2021epo,Reichert:2021cvs,Morgante:2022zvc,He:2022amv,Fujikura:2023fbi,Pasechnik:2023hwv}.

In the case of the ultrarelativistic regime, one may use the ballistic approximation~\cite{Dine:1992wr,Moore:1995si}. Mathematically, the ballistic approximation amounts to 
\bea 
\label{Eq:collision_less}
L_w \ll \gamma_w L_{\rm MFP} \qquad \text{(ballistic condition)},
\eea 
where $L_w$ is the width of the wall in the wall frame, $L_{\rm MFP}$ is the mean free path of the incoming particles in the plasma frame. In this case, the particles do not (have time to) collide with each other when they pass through the wall. The pressure on the bubble wall is simply due to the transmission of the particles from the outside to the inside, and their reflection from the wall. It is generally a \emph{monotonically increasing} function of the velocity. This pressure, at large $\gamma_w \gg 1$, asymptotes to~\cite{Bodeker:2009qy}
\begin{align}
\label{eq:BM-force}
    \mathcal{P}_{\rm BM}\simeq \sum_i g_i c_i \frac{\Delta m^2_i  T_n^2}{24}\,,
\end{align}
where $c_i=1(1/2)$ for bosons (fermions), $g_i$ is the number of internal degrees of freedom, $T_n$ is the nucleation temperature, and $\Delta m_i$ is the mass gain of the particle as it transits from the exterior to the interior of the bubble. The criterion 
\bea 
\label{eq:BMcriterion}
\P_{\rm driving}> \P_{\rm BM} \qquad \text{(B\"{o}deker and Moore (BM) criterion)},
\eea 
where $\P_{\rm driving}$ is the driving pressure, is frequently used to determine whether a wall runs away or not. Note that $\P_{\rm BM}$ in Eq.~\eqref{eq:BM-force} is only the friction in the ultrarelativistic regime at the leading order in $\gamma_w$. Other particle-number-changing processes typically induce additional friction that depends on $\gamma_w$ \cite{Bodeker:2017cim,Hoche:2020ysm,Gouttenoire:2021kjv,Ai:2023suz}. There can also be additional friction due to heavy new physics~\cite{Azatov:2020ufh} that is also independent of $\gamma_w$. In this paper, we will not consider those types of contributions. Therefore, one needs to assume that there is no additional $\gamma_w$-independent friction in the ultrarelativistic regime other than the BM thermal friction. And we interpret ``runaway" in this paper as that the wall can enter into a regime with extremely large $\gamma_w$ for which other $\gamma_w$-dependent (if present) pressures become comparable with $\P_{\rm BM}$.

However, the total friction on the wall does not need to be a monotonous function of the wall velocity and could have a maximum before it reaches the BM limit. Indeed, it is observed in Refs.~\cite{Cline:2021iff,Laurent:2022jrs,Ai:2023see} that there is a pressure barrier peaked at the Jouguet velocity $\xi_J$. This peak is largely due to the hydrodynamic obstruction~\cite{Konstandin:2010dm}, a frictional force that exists in LTE, as shown in Ref.~\cite{Laurent:2022jrs}.\footnote{Ref.~\cite{Dorsch:2023tss} argues that out-of-equilibrium effects should be comparable to the hydrodynamic obstruction, based on the so-called extended fluid Ansatz~\cite{Dorsch:2021nje,Dorsch:2021ubz}.} The hydrodynamic obstruction will also be called the LTE force occasionally.

In this paper, we compare the maximal hydrodynamic obstruction with the BM thermal friction Eq~\eqref{eq:BM-force}. We show that the former can be larger than the latter in large regions of the parameter space and hence the runaway criterion must be reconsidered. This means that in many models while the BM thermal friction cannot stop the wall's acceleration, the hydrodynamic obstruction can however do this job. We work out the critical phase transition strengths, $\alpha_{n,\rm crit}^{\rm BM/LTE}$, above which the driving force cannot be balanced out by the maximal hydrodynamic obstruction/the BM thermal friction, and compare them with each other. Defining the ratio of the number of degrees of freedom (DoFs) in the broken and symmetric phases, $b \equiv a_-/a_+$, we find that hydrodynamic obstruction becomes more relevant than the BM thermal friction in prohibiting runaway for $b \lesssim 0.85$. Although the value of $b$ in the standard model (SM) at the electroweak phase transition (EWPT) is close to one, $b_{\rm EWPT} \sim 0.9$, many beyond-standard-model (BSM) phase transitions feature a much smaller $b$, like in minimal toy models of phase transition with two scalar fields~\cite{Azatov:2019png}, minimal models of $B-L$ breaking~\cite{Jinno:2016knw,Chun:2023ezg}, Peccei-Quinn symmetry breaking~\cite{DelleRose:2019pgi}, or conformal symmetry breaking~\cite{Prokopec:2018tnq,Jinno:2016knw}. Deconfinement-confinement phase transitions like in the composite Higgs also feature a large change in the number of DoFs~\cite{DeCurtis:2019rxl,Baldes:2021aph,Sanchez-Garitaonandia:2023zqz, Azatov:2020nbe}. 

The maximal hydrodynamic obstruction has been partly studied in Ref.~\cite{Ai:2023see}. There, however, no attention has been paid to the comparison with the BM thermal friction. Furthermore, we give a simpler derivation of the critical phase transition strength $\alpha_{n,\rm crit}^{\rm LTE}$ without solving the fluid equations. Therefore, our derivation can be very easily implemented in, e.g., Mathematica. We also provide a fit for the obtained $\alpha_{n,\rm crit}^{\rm LTE}(b)$ as a function of $b$.

The paper is organized as follows. In the next section, we derive the general formulae for the driving and frictional pressures for general wall motions (not necessarily stationary) from the EoM of the scalar field. In Sec.~\ref{sec:match-conditions}, we derive a different formula for the frictional pressure based on the matching conditions. In Sec.~\ref{sec:maxial_LTE_pressure_det}, we study the maximal LTE friction {\it in the detonation regime} which however is {\it not} the maximal hydrodynamic obstruction. The latter occurs in the hybrid regime and is studied in Sec.~\ref{sec:maximal_hydrodynamic_obstruction}. In Sec.~\ref{sec:compare_with_BM}, we compare the maximal hydrodynamic obstruction with the BM thermal friction. In Sec.~\ref{sec:comment_on_LTE}, we comment on the validity of the approximations used in this work and give some caveats. We conclude in Sec.~\ref{sec:conc}. In the Appendix, we also briefly discuss how to include the out-of-equilibrium effects in the analysis.

\section{Bubble wall dynamics and the friction}
\label{Sec:bubble_dyn}
Analysis of the friction on bubble walls is usually based on the following EoM for the background field~\cite{Moore:1995ua,Moore:1995si},
 \begin{align}
 \label{eq:eom}
  \Box\phi+\frac{\d V(\phi)}{\d\phi}+\sum_i\frac{\d m^2_i(\phi)}{\d\phi}\int \frac{\d^3{\bf p}}{(2\pi)^32E_i}\,f_i(p,x)=0,
 \end{align}
where $\Box=\partial_\mu \partial^\mu$,  $f_i(p,x)$ are the particle distribution functions, and $E_i=\sqrt{\vec{p}^2_i+m_i^2}$ the particle energies. Here $x$ and $p$ denote general four-dimensional position and momentum coordinates. Below, without loss of generality, we shall assume a planar wall expanding in the $z$ direction.

For a bubble wall with a constant velocity, which we will refer to as a \emph{stationary} wall, one can always work in the wall rest frame where all quantities are time independent. However, for a general motion, we cannot define a global rest frame of the wall. But we can still define an ``instantaneous'' wall rest frame which can help us find the expressions for the driving and frictional pressures. With fixed $x$ (to be distinguished from the general four-dimensional coordinate) and $y$, the worldline of the wall can be parameterized by a proper time $\tau$. Near an arbitrarily chosen proper time of the wall, we define an instantaneous comoving rest frame as follows. First, we assume a certain Ansatz for the wall in the plasma frame from which the wall velocity and acceleration can be read~\cite{Ai:2022kqm}. For example, one can take ($v_b$ is the symmetry broken value of $\phi$)
\begin{align}
\label{eq:wall-profile}
    \phi(t',z')=\frac{v_b}{2}\big(1-{\rm tanh} [(z'-z'_0(t'))/L'_w]\big),
\end{align}
where $\{t',z'\}$ are the coordinates in the plasma frame and $L'_w$ is the wall width observed in that frame. The function $z'_0(t')$ describes the trajectory of the wall. We can choose $t'=0$ at the reference point of the worldline (corresponding to the chosen $\tau$). Then $z'_0(t)=z'_0+v_w t'+\frac{1}{2} a t'^2+\O(t'^3)$. The velocity $v_w$ is used to define the (instantaneous) comoving coordinates 
\begin{align}
    z=\gamma_w(z'-z'_0-v_w t'),\quad t=\gamma_w[t'-v_w (z'-z'_0)].
\end{align}
In the instantaneous rest frame of the wall, one can choose $z=0$ as the position of the wall and at the origin of coordinates $t=0$. It is easy to show that the first time derivative of $\phi$ in the instantaneous comoving frame, $\partial_t{\phi}(0)\equiv \dot{\phi}(0)$, is vanishing. The second time derivative of $\phi$, $\ddot{\phi}(0)$, depends on the acceleration $a$ and vanishes also if $a=0$ which corresponds to a stationary wall. 

In the instantaneous comoving frame, multiplying the EoM  by $\partial_z\phi$ and integrating over $z$, one gets
\begin{align}
  \int_{-\delta}^\delta \d z\,(\partial_z\phi)\left(\ddot{\phi}(0)|_{\tau}-\partial_z^2\phi+\frac{\d V(\phi)}{\d\phi}+\sum_i\frac{\d m^2_i(\phi)}{\d\phi}\int \frac{\d^3{\bf p}}{(2\pi)^32E_i}\,f_i(p,z)\right)=0,
\end{align}
where $\delta$ is a length much larger than the wall width, and effectively can be taken to be infinity in the present context. Above, we also have Taylor expanded quantities in $t$ at $t=0$ and kept the leading contributions. The second term is vanishing because the integrand $(\partial_z\phi) \partial^2_z\phi$ is a total derivative and $\partial_z\phi$ vanishes for large $|z|$ (for any $\tau$). The term with $\ddot{\phi}(0)|_{\tau}$ is related to the acceleration of the wall at $\tau$ as discussed above. Thus we obtain
\begin{align}
\label{eq:balance}
    -\int_{-\delta}^\delta \d z\, (\partial_z\phi)\, \ddot{\phi}(0)|_{\tau} = \P_{\rm driving}-\P_{\rm friction},
\end{align}
where 
\begin{align}
\label{eq:driving}
    \P_{\rm driving}=\int_{-\delta}^\delta \d z\,(\partial_z\phi) \frac{\d V(\phi)}{\d\phi}=\int_{-\delta}^\delta \d z\, \frac{\d V(\phi(z))}{\d z},
\end{align}
and 
\begin{align}
\label{eq:friction-general}
    \P_{\rm friction}= - \int_{-\delta}^\delta \d z\, (\partial_z\phi)\left(\sum_i\frac{\d m^2_i(\phi)}{\d\phi}\int \frac{\d^3{\bf p}}{(2\pi)^32E_i}\,f_i(p,z)\right).
\end{align}
The sum in the last expression is to be taken on the species changing mass across the wall. 
Although the above expressions implicitly depend on $\tau$, we expect that their dependence on $\tau$ from the $\tau$-dependence in $\phi$ disappears or becomes very weak as long as $\delta$ is much larger than the wall width. Thus $\P_{\rm driving}$ can be identified as the zero-temperature potential difference between the symmetric and broken phases,
\begin{align}
\label{eq:driving2}
    \P_{\rm driving}=\Delta V,
\end{align}
and is therefore not a function of the temperature nor $\tau$. $\P_{\rm friction}$ can still be $\tau$-dependent due to the time-dependence in the distribution functions which should be determined by solving the Boltzmann equation. A stationary motion is reached when $\P_{\rm driving}=\P_{\rm friction}$, which leads to  $\ddot{\phi}(0)|_{\tau}=0$, as expected. Looking at Eq.~\eqref{eq:friction-general}, it would seem that $\P_{\rm friction}$ only depends on bath particles directly coupling with the scalar field $\phi$ with a wall-dependent mass (particles that we will call \emph{active particles} in what follows). This is not exactly the case, because the distribution functions depend on the interactions among all the particles in the bath. Writing $f_i=f_i^{\rm eq}+\delta f_i$, the frictional pressure can be further decomposed into two parts
\begin{align}
\label{eq:Pfriction_as_a_sum}
    \P_{\rm friction}=\P_{\rm LTE}+\P_{\rm dissipative},
\end{align}
where $\P_{\rm dissipative}$ is given by Eq.~\eqref{eq:friction-general} with $f_i$ replaced by the out-of-equilibrium part $\delta f_i$. In this paper, we are interested in $\P_{\rm LTE}$ which reads
\begin{align}
    \P_{\rm LTE}&=- \int_{-\delta}^\delta \d z\, (\partial_z\phi)\left(\sum_i\frac{\d m^2_i(\phi)}{\d\phi}\int \frac{\d^3{\bf p}}{(2\pi)^32E_i}\,f^{\rm eq}_i(p,z;T)\right)\notag\\
    &=-\int_{-\delta}^\delta\d z\, (\partial_z\phi) \frac{\partial V_T(\phi,T)}{\partial \phi}.
\end{align}
Here $V_T(\phi,T)$ is the finite-temperature corrections to the potential in the non-interacting-gas approximation~\cite{Espinosa:2010hh} and $T$, which in general depends on $z$, is the temperature of all the particles being in local thermal equilibrium. Remembering the $z$-dependence in $V_T(\phi(z), T(z))$ and using the chain rule for derivatives, $\P_{\rm LTE}$ can be further written as
\begin{align}
\label{eq:PLTE1}
    \P_{\rm LTE}=-\Delta V_T+ \int_{-\delta}^\delta\d z\, \frac{\partial V_T}{\partial T}\frac{\partial T}{\partial z}.
\end{align}
Note that there is no clear criterion to distinguish between the driving and friction forces. Different authors may attribute the same force to either the driving force or the friction force. Our definitions above are different from those used in Ref.~\cite{Ai:2021kak} where $\Delta V_T$ was identified as a part of the driving force.\footnote{With this identification, the friction is contributed only from the second term in Eq.~\eqref{eq:PLTE1} and is non-vanishing only when $T$ is inhomogeneous. } However, the quantity $\P_{\rm friction}-\P_{\rm driving}$  is invariant under different identifications. A merit of the definitions we are using is that the driving pressure is constant, in agreement with the convention used in Refs.~\cite{Bodeker:2009qy,Bodeker:2017cim}. 

In this paper, we shall assume LTE for the plasma and study $(\P_{\rm friction}-\P_{\rm driving})(\xi_w)$ as a function of the wall velocity. In particular, we are interested in its maximum, $(\P_{\rm LTE}-\P_{\rm driving})_{\rm max} \equiv (\P_{\rm friction}-\P_{\rm driving})_{\xi_w = \xi_{\rm peak}}$ (see the left panel of Fig.~\ref{fig:total-pressure} below). If $(\P_{\rm LTE}-\P_{\rm driving})_{\rm max}>0$, then there should be a stationary solution for the wall in LTE. On the other hand, if $(\P_{\rm LTE}-\P_{\rm driving})_{\rm max}<0$, a stationary solution is not possible and the wall runs away unless $\P_{\rm dissipative}$ becomes large enough to stop the acceleration of the wall. Since $\P_{\rm driving}$ is constant, $(\P_{\rm LTE}-\P_{\rm driving})_{\rm max}$ corresponds to $\P_{\rm LTE}^{\rm max}$, the maximal LTE frictional pressure.

In realistic situations, the plasma is typically out of equilibrium. However, if we assume that out-of-equilibrium effects would add more friction to the wall, which is confirmed in a simple model in Ref.~\cite{Laurent:2022jrs} and also in the Appendix for general case, $\P^{\rm max}_{\rm LTE}>\P_{\rm driving}$ implies $\P^{\rm max}_{\rm friction}>\P_{\rm driving}$ and the former can serve as a conservative condition for non-runaway bubble walls. 

It is still not clear how to compute $\P_{\rm LTE}$ using the formula~\eqref{eq:PLTE1} directly. In the next section, we derive another form of $\P_{\rm LTE}$  that can be conveniently used to determine $\P_{\rm LTE}^{\rm max}$.

\section{Matching conditions for stationary and non-stationary walls}
\label{sec:match-conditions}

The hydrodynamics of the bubble growth is based on the total energy-momentum conservation. Usually, a stationary wall is assumed and one works in the rest frame of the wall. Here again, we do not assume a stationary motion of the wall. The energy-momentum tensors for the scalar background and the fluid read
\begin{subequations}
\begin{align}
T_{\phi}^{\mu\nu}&=(\partial^\mu\phi)\partial^\nu\phi-g^{\mu\nu}\left(\frac{1}{2}(\partial\phi)^2-V(\phi)\right),\\
T_f^{\mu\nu}&=(e_f+p_f)u^\mu u^\nu- g^{\mu\nu}p_f,
\end{align}
\end{subequations}
where $u^\mu$ is the fluid four-velocity, {$e_f$} and $p_f$ are the fluid energy density and pressure that vanish at zero temperature.  However, one usually combines the fluid energy density and pressure with the tree-level scalar potential energy, $e=e_f+V(\phi)$, $p=p_f-V(\phi)$. The advantage of using them is that in terms of $e$ and $p$ the matching conditions for hydrodynamic quantities take the standard form that appears commonly in the literature. Note that the fluid enthalpy $\omega=e_f+p_f=e+p$. In terms of $e$ and $p$, the energy-momentum tensor for the fluid then takes the form as
\begin{align}
    T_f^{\mu\nu}&=(e+p)u^\mu u^\nu- g^{\mu\nu}[p+V(\phi)].
\end{align}

\begin{figure}
    \centering
    \includegraphics[scale=0.5]{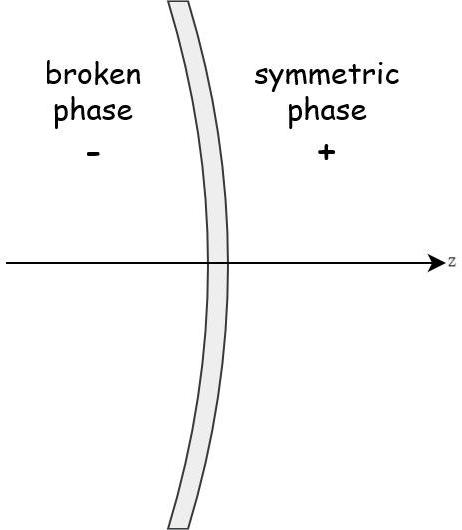}
    \caption{Rest frame of the bubble wall used for the matching conditions.}
    \label{fig:wall}
\end{figure}

The matching conditions are obtained from the $\nu=0$ and $\nu=3$ components of  $\nabla_\mu T^{\mu\nu}=0$. Working in the instantaneous comoving frame and writing $u^{\mu}=\gamma(z) (1,0,0,-v(z))$, one obtains
\begin{subequations}
\label{eq:matching-original}
    \begin{align}
         &\ \partial_z (\omega \gamma^2 v) =0,\\
        &-(\partial_z\phi)\, \ddot{\phi}(0)|_{\tau}+\partial_z \left[\omega \gamma^2 v^2+\frac{1}{2}(\partial_z\phi)^2+p\right] =0,
    \end{align}
\end{subequations}
where $\omega\equiv (e+p)$ is the enthalpy. Above, we have assumed that all the first derivatives in time of the hydrodynamic quantities are vanishing.\footnote{Whether this assumption is true or not cannot be solely determined by the total energy-momentum conservation. It depends on the microscopic details of how the scalar field interacts with the plasma. If this assumption is not true, one would have additional terms in Eqs.~\eqref{eq:matching-original} expressed by the first derivatives in time of the hydrodynamic quantities. But since for $\ddot{\phi}(0)|_\tau=0$ (i.e. vanishing acceleration of the wall) the time dependence disappears in all quantities in the comoving frame, the first time derivatives of the hydrodynamic quantities in the instantaneous comoving frame must depend on the acceleration of the wall $a$ and vanish as $a\rightarrow 0$. Since the critical phase transition strength $\alpha_{n,\rm crit}^{\rm LTE}$ is given ultimately by a stationary motion ($a=0$, see below), our derivation of it would not be changed even if the present assumption is not fully true.} Integrating over $z$, one obtains
\begin{subequations}
\label{eq:junctionAB}
\begin{align}
    \omega_+\gamma_+^2v_+ &=\omega_-\gamma_-^2v_-,\label{eq:conditionA}\\
     -\int_{-\delta}^\delta \d z\, (\partial_z\phi)\, \ddot{\phi}(0)|_{\tau} &=[\omega_-\gamma_-^2v_-^2+p_-]-[\omega_+\gamma_+^2v_+^2+p_+] \notag\\
    &\equiv \Delta V- \Delta(-V_T+\omega \gamma^2 v^2),\label{eq:conditionB}
\end{align}
\end{subequations}
where a subscript ``$\pm$'' denotes quantities in front of/behind the bubble wall. To be explicit, $\omega_+=\omega_s(T_+)$, $\omega_-=\omega_b(T_-)$ (and similarly for $p_{\pm }$), where the label ``$s/b$'' denotes the symmetric/broken phase, see Fig.~\ref{fig:wall}. In the last equality of Eq.~\eqref{eq:conditionB} we have used the thermodynamic relation in LTE $p=-V_{\rm eff}(\phi) \equiv - V(\phi) - V_T(\phi)$, and so it is valid only for LTE~\cite{Ai:2021kak}. Comparing Eq.~\eqref{eq:conditionB} with Eqs.~\eqref{eq:balance} and~\eqref{eq:driving2}, we thus obtain
\begin{align}
\label{eq:P-LTE-def}
    \P_{\rm LTE}= -\Delta V_T+\overbar{\P}_{\rm LTE},
\end{align}
where
\begin{align}
\label{eq:Pbar-LTE-def}
    \overbar{\P}_{\rm LTE}\equiv \Delta  \{\omega \gamma^2 v^2\}  =  \Delta \{ (\gamma^2-1)Ts\}
\end{align}
with $s=\omega/T$ being the entropy density. Using Eq.~\eqref{eq:conditionA}, one can rewrite it as
\begin{align}
\label{eq:LTE-pressure2}
    \overbar{\P}_{\rm LTE}= \omega_+ \gamma_+^2 v_+ (v_+-v_-).
\end{align}

In LTE, there is an additional matching condition due to the entropy conservation~\cite{Ai:2021kak}\footnote{The derivation of this condition involves only first derivatives in time and hence should be valid in the instantaneous comoving frame for non-stationary walls.}
\begin{equation} 
\label{eq:LTE-matching_1}
   \partial_\mu (s u^\mu) = 0\quad\Rightarrow\quad  s_+ \gamma_+ v_+ =s_- \gamma_- v_-.
\end{equation}
The above condition is a special situation of the non-negativity of entropy production~\cite{Laine:1993ey}.
Using Eq.~\eqref{eq:conditionA}, it can be also written as
\begin{align}
 \label{eq:LTE-matching}
\frac{T_+}{T_-}=\frac{\gamma_-}{\gamma_+}\,.
\end{align}
In Appendix \ref{App:entropy_inj}, we explain how to extend this picture if there is a large entropy injection and we argue that the LTE limit provides a \emph{lower bound} on the resistance to the bubble expansion. 

If we impose $\ddot{\phi}(0)|_\tau=0$, i.e., vanishing acceleration, we arrive at three matching conditions for three unknown hydrodynamic quantities $\{T_-,v_+, v_-\}$ ($T_+$ can be related to the nucleation temperature $T_n$ which is a given parameter characterizing the phase transition), i.e. Eqs.~\eqref{eq:junctionAB} (with $\ddot{\phi}(0)|_\tau=0$) and Eq.~\eqref{eq:LTE-matching}. We therefore can fully determine all these three quantities, and thus also the wall velocity $\xi_w$ (e.g., for detonations, $\xi_w=v_+$). To study the frictional force in LTE as a function of $\xi_w$, $\ddot{\phi}(0)$ must be allowed to take any value so that the wall velocity $\xi_w$ becomes a variable. In this case, one should in principle only use the matching conditions~\eqref{eq:conditionA},~\eqref{eq:LTE-matching} and the definition of $\P_{\rm LTE}$ given in Eqs~\eqref{eq:P-LTE-def},~\eqref{eq:Pbar-LTE-def}. However, as we will see shortly, for the derivation of the critical phase transition strength $\alpha_{n,\rm crit}^{\rm LTE}$, $\Ddot{\phi}(0)|_\tau=0$ can be imposed.

It is argued in Ref.~\cite{Ai:2023see}, based on numerical results and previous works~\cite{Cline:2021iff, Laurent:2022jrs}, that the hydrodynamic obstruction is maximal when the wall velocity $\xi_w$ is very close to the Jouguet velocity $\xi_J$. We will elaborate on this point in Sec. \ref{sec:maximal_hydrodynamic_obstruction}. We can however already anticipate the important points: the hydrodynamic obstruction originates mainly from the heating due to the shock wave existing in front of the bubble, which decreases the effective phase transition strength felt by the bubble wall. This heating becomes more efficient as the velocity of the shock-wave front
increases $\xi_{\rm sw}$ (see Eq.~\eqref{eq:sw-matching1} and the heating parameter Eq.~\eqref{eq:heating}). The velocity of the shock-wave front
has however to be smaller than the Jouguet velocity $\xi_{\rm sw} \leq \xi_{J}$, which implies that the maximal heating, and thus obstruction, will occur at the Jouguet velocity. For larger wall velocities, the LTE pressure decreases (when the motion becomes a detonation), because the shock wave disappears, and thus the heating vanishes.   

\begin{figure}[ht]
    \centering
\includegraphics[width=0.45\linewidth]{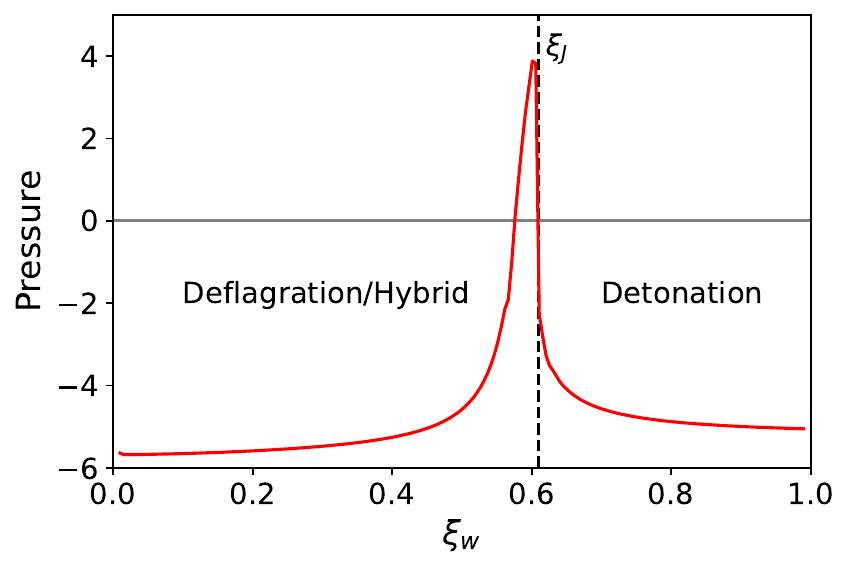}
\includegraphics[width=0.45\linewidth]
{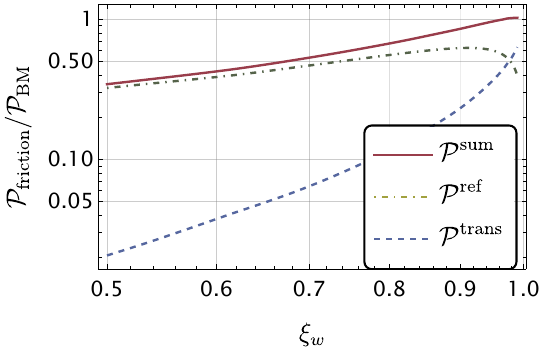}
    \caption{ Left: Example of $\P_{\rm LTE}-\P_{\rm driving}$ (in arbitrary units) as a function of the wall velocity $\xi_w$ for a given phase transition strength $\alpha_n$, based on results of Ref.~\cite{Laurent:2022jrs}. $\P_{\rm LTE}^{\rm max}$ occurs slightly before the Jouguet velocity $\xi_J$. Plot taken from Ref.~\cite{Ai:2023see}.  Right: Example of the total frictional pressure due to transmission and reflection in the ballistic picture that are assumed to become more and more exact for a larger and larger $\gamma_w$. It is a monotonically increasing function and saturates at the BM value.
    } 
    \label{fig:total-pressure}
\end{figure}

The typical behaviour of the total pressure on the wall in LTE is shown in the left panel of Fig.~\ref{fig:total-pressure}. Since the Jouguet velocity is a connecting point between the hybrid and detonation regimes, one might think that $\P^{\rm max}_{\rm LTE, hyb}=\P^{\rm max}_{\rm LTE, det}$. However, this is generically not true as indicated by the very steep slope at $\xi_J$. The reason behind this is the sudden disappearance of the shock wave when one moves from the hybrid regime to the detonation regime which typically leads to a lower $\P^{\rm max}_{\rm LTE, det}$ compared with $\P^{\rm max}_{\rm LTE, hyb}$. Only when the ratio of the enthalpies in the broken and
symmetric phases is very close to one, is the equality  $\P^{\rm max}_{\rm LTE, hyb}=\P^{\rm max}_{\rm LTE, det}$ satisfied~\cite{Ai:2023see}, see Sec.~\ref{sec:maximal_hydrodynamic_obstruction} for more details. Therefore, we have 
\begin{align}
    \P^{\rm max}_{\rm LTE}=\P^{\rm max}_{\rm LTE, hyb} \geq \P_{\rm LTE,det}^{\rm max}.
\end{align}

As $\alpha_n$ increases, one should expect that the peak of the red curve in the left panel of Fig.~\ref{fig:total-pressure} would move down due to the increase of $\P_{\rm driving}$. At a critical value, denoted as $\alpha_{n,\rm crit}^{\rm LTE}$, the peak would touch exactly the zero line. For bigger $\alpha_n$, $\P_{\rm LTE}-\P_{\rm driving}$ cannot equal zero for all $\xi_w\in [0,1)$ and thus the bubble wall runs away assuming LTE. Therefore, this value $\alpha_{n,\rm crit}^{\rm LTE}$ provides a criterion for runaway walls in LTE;
\begin{align}
\text{LTE runaway criterion}: \alpha_n > \alpha_{n,\rm crit}^{\rm LTE}. 
\end{align}

Note that the peak has a narrow width,  which means that when $\alpha_n <\alpha_{n,\rm crit}^{\rm LTE}$, the wall velocity is typically close to $\xi_J$ such that $\xi_w\approx \xi_J$ may be a good approximation when the stationary motion is caused by hydrodynamic obstruction~\cite{Laurent:2022jrs}. Since the peak decreases dramatically after $\xi_J$, only deflagration/hybrid (or ultrarelativistic detonation) solutions are viable~\cite{Ai:2023see}. This may also be true if out-of-equilibrium effects are included~\cite{Laurent:2022jrs} (see, however, Ref.~\cite{Krajewski:2023clt}). 

Similarly, there is also the runaway criterion based on the BM thermal friction on ultrarelativistic bubble walls~\cite{Bodeker:2009qy}; 
\begin{align}
    \text{B\"{o}deker-Moore runaway criterion}: \alpha_n > \alpha_{n,\rm crit}^{\rm BM}.
\end{align}
In the right panel of Fig.~\ref{fig:total-pressure}, we show the frictional pressures in the ballistic approximation, obtained by computing the momentum exchange between the wall and particles that transmit from the outside to the inside of the wall $\P^{\rm trans}$, and that bounce off from the wall $\P^{\rm reflection}$. In this type of calculation, the plasma outside of the wall is assumed to be unperturbed, and thus the calculation is typically valid for a relativistic detonation regime. The total frictional pressure $\P^{\rm sum}$ approaches the BM limit for $\gamma_w\gg 1$ where $\P^{\rm reflection}$ approaches zero. It is expected that the red curve in the left panel of Fig.~\ref{fig:total-pressure} is followed by the dark red curve (when $-\P_{\rm driving}$ is included) from the right panel when the same particle physics model is considered. Note that, however, in some specific models, one may still have a non-monotonic behaviour of the pressure as a function of $\xi_w$ (or $\gamma_w$) even in the relativistic regime. In Ref.~\cite{GarciaGarcia:2022yqb}, it was found that massive dark photons coupling with the order-parameter scalar can induce a non-monotonic dependence on $\gamma_w$ for the friction. The friction then has a peak at intermediate $\gamma_w$ that the authors dubbed maximum dynamic pressure. We however do not consider such a situation and assume that such an additional peak is absent.

Since $\alpha_{n,\rm crit}^{\rm LTE}$ and $\alpha_{n,\rm crit}^{\rm BM}$ are not necessarily equal to each other, we propose a modified runaway criterion in this work;
\begin{align}
    {\rm Modified\ runaway\ criterion}: \alpha_n> {\rm max}\{\alpha_{n,\rm crit}^{\rm LTE}, \alpha_{n,\rm crit}^{\rm BM}\}.
\end{align}

Below, we shall derive both $\alpha_{n,\rm crit}^{\rm LTE}$ and $\alpha_{n,\rm crit}^{\rm BM}$, and compare them. As we shall see, $\alpha_{n,\rm crit}^{\rm LTE}$ can be larger than $\alpha_{n,\rm crit}^{\rm BM}$ in large parameter space, indicating that the peak of the LTE force is larger than the BM thermal friction. For, $\alpha_n\in (\alpha_{n,\rm crit}^{\rm BM},\alpha_{n,\rm crit}^{\rm LTE}]$, the wall acceleration would be stopped by the LTE force even though it cannot be stopped by the BM thermal friction that exists in the ultrarelativistic regime. 

Since $\alpha_{n,\rm crit}^{\rm LTE}$ is given by $\P_{\rm LTE}^{\rm max}=\P_{\rm driving}$, i.e., when the peak touches the zero line of Fig.~\ref{fig:total-pressure} (left panel), we actually can use the additional matching condition (that is valid only for stationary walls)
\begin{align}
\label{eq:conditionB2}
    \omega_-\gamma_-^2v_-^2+p_- =\omega_+\gamma_+^2v_+^2+p_+,
\end{align}
in deriving $\alpha_{n,\rm crit}^{\rm LTE}$. With the conditions~\eqref{eq:conditionA} and~\eqref{eq:conditionB2}, one has~\cite{Ai:2021kak}
\begin{align}
\label{eq:v-1}
    v_-=\frac{1}{6v_+}\left[1-3 \alpha _++3 \left(1+\alpha _+\right) v_+^2\pm\sqrt{\left(1-3 \alpha _++3 \left(1+\alpha _+\right) v_+^2\right){}^2-12 v_+^2}\right],
\end{align}
where
\begin{align}
\label{eq:alpha&r}
    \alpha_+\equiv \frac{\Delta\theta}{3\omega_+},\quad {\rm with}\quad \theta=\rho-3p,
\end{align}
and $+$ and $-$ sectors correspond to deflagration/hybrid and detonation regimes, respectively. 
One can then substitute Eq.~\eqref{eq:v-1} into Eq. \eqref{eq:LTE-pressure2}.

\section{Maximal LTE pressure in the detonation regime}
\label{sec:maxial_LTE_pressure_det}

Before we discuss the maximal hydrodynamic obstruction in LTE, equivalently, $\alpha_{n,\rm crit}^{\rm LTE}$, we first discuss the maximal LTE pressure in the {\it detonation} regime, $\alpha_{n,\rm max}^{\rm det}$. This simpler analysis serves as a warm-up for the more complicated analysis given in the next section. Below, we carry out our analysis for the bag equation of state (EoS) and one of its generalizations where the sound speed can deviate from $1/\sqrt{3}$.

\subsection{Bag model}
To close the system of equations, we now need to introduce an EoS. We first consider the \emph{Bag model} for which we have
\begin{align}
\label{eq:bag_eos}
    &e_s(T)=a_+  T^4+\epsilon,\qquad p_s(T)=\frac{1}{3}a_+ T^4-\epsilon,\\
    &e_b(T)=a_- T^4 ,\ \quad\qquad\  p_b(T)=\frac{1}{3}a_- T^4,
\end{align}
where $a_\pm$ and $\epsilon$ are constants and we used the convention $\epsilon \equiv \Delta V$. With this EoS, $\alpha_+ = \epsilon/(a_+ T_+^4)$.

In the detonation regime, $v_+=\xi_w$, $T_+=T_n$, $\alpha_+=\alpha_n\equiv\epsilon/(a_+ T_n^4)$. Substituting these relations into Eqs.~\eqref{eq:v-1} and~\eqref{eq:LTE-pressure2}, one obtains~\cite{Ai:2021kak} 
\begin{align}
\label{eq:Pbar-LTE}
    \overbar{\P}_{\rm LTE}=a_+ T_n^4 f(\xi_w,\alpha_n),
\end{align}
where 
\begin{align}
\label{eq:f-function}
    f(\xi_w,\alpha_n)=-\frac{2\gamma_w^2}{9} \left[1-3\alpha_n+3(\alpha_n-1)\xi_w^2+\sqrt{\left[1-3\alpha_n+3(1+\alpha_n)\xi_w^2\right]^2-12 \xi_w^2} \right].
\end{align}
Since in deriving $f(\xi_w,\alpha_n)$, one has used the matching condition~\eqref{eq:conditionB2}, for given $\alpha_n$ this expression is valid only for the particular values of $\xi_w$ that satisfy $\P_{\rm LTE}=\P_{\rm driving}$. Alternatively, one can view $\xi_w$ as a free variable but $\alpha_n$ as an implicit function of $\xi_w$ defined by $\P_{\rm LTE}=\P_{\rm driving}$. From this perspective, $\alpha_n(\xi_J)$ defines $\alpha_{n,\rm max}^{\rm det}$.

The Jouguet velocity $\xi_J$ corresponds to $v_-=1/\sqrt{3}$~\cite{Espinosa:2010hh}
\begin{align}
\label{eq:Jouguet-velocity-bag}
    \xi_J=\frac{1}{\sqrt{3}}\left(\frac{1+\sqrt{3\alpha_n\left(\frac{2}{3}+\alpha_n\right)}}{1+\alpha_n}\right).
\end{align}
At $\xi_w=\xi_J$, the square-root term in Eq.~\eqref{eq:f-function} is vanishing. Substituting $\xi_w=\xi_J$ into $f(\xi_w,\alpha_n)$, one obtains
\begin{align}
\label{eq:fmax}
    f(\alpha_{n,\rm max}^{\rm det})=\frac{2}{3}\frac{3(\alpha_{n,\rm max}^{\rm det})^2+\alpha_{n,\rm max}^{\rm det}+(1-\alpha_{n,\rm max}^{\rm det})\sqrt{3(\alpha_{n,\rm max}^{\rm det})^2+2\alpha_{n,\rm max}^{\rm det} }}{2\alpha_{n,\rm max}^{\rm det}+1-\sqrt{3(\alpha_{n,\rm max}^{\rm det})^2+2\alpha_{n,\rm max}^{\rm det}}},
\end{align}
where we have replaced $\alpha_n$ with $\alpha_{n,\rm max}^{\rm det}$. One can approximate $f(\alpha_{n,\rm max}^{\rm det})$ with a linear function 
\bea 
\label{eq:liear-approx}
    f(\alpha_{n,\rm max}^{\rm det}) \approx 0.3 + 3.155 \alpha_{n,\rm max}^{\rm det}.
\eea 
Substituting Eq.~\eqref{eq:liear-approx} into Eq.~\eqref{eq:Pbar-LTE}, we obtain
\begin{align}
    \overbar{\P}^{\rm max}_{\rm LTE,det}(\alpha_{n,\rm max}^{\rm det})\approx a_+ T_n^4 (0.3+3.155 \alpha_{n,\rm max}^{\rm det}).
\end{align}
This approximated expression should work well for $\alpha_{n,\rm max}^{\rm det} \gtrsim 0.2$. For very small $\alpha_{n,\rm max}^{\rm det}$, one should use the exact result obtained from Eq.~\eqref{eq:fmax}.

For $-\Delta V_T$, we have
\begin{align}
    -\Delta V_T= \frac{a_+T_n^4}{3} \left(1 -b r^4 \right),
\end{align}
where $r \equiv \gamma_w/\gamma_-$ and $b \equiv a_-/a_+$. Above, we have used the LTE matching condition~\eqref{eq:LTE-matching}. Substituting $v_-=1/\sqrt{3}$, $\xi_w=\xi_J$ into $r$, one obtains
\begin{align}
    r^4_J(\alpha_n)\equiv \left. r^4(\alpha_n)\right|_{v_-=1/\sqrt{3},\,\xi_w=\xi_J}=\frac{(\alpha_n+1)^4}{[2\alpha_n+1-\sqrt{\alpha_n(3\alpha_n+2)}]^2} \overset{\alpha_n \to \infty}{\longrightarrow} (7+4\sqrt{3})\alpha_n^2.
\end{align}
Therefore, $-\Delta V_T$ evaluated at $\xi_w=\xi_J$ is 
\bea 
(-\Delta V_T)|_{\xi_J} (\alpha_n) =a_+ T_n^4\left(\frac{1}{3} - \frac{b(\alpha_n+1)^4}{3 [2\alpha_n+1-\sqrt{\alpha_n(3\alpha_n+2)}]^2}\right) \overset{\alpha_n \to \infty}{\longrightarrow} -   \frac{b(7+4\sqrt{3})\alpha_n^2}{3} a_+ T_n^4 .
\eea 

Finally, the driving pressure is 
\begin{align}
    \P_{\rm driving}(\alpha_n) =a_+ T_n^4 \alpha_n.
\end{align}

The equation $-\Delta V_T|_{\xi_J}(\alpha_{n,\rm max}^{\rm det})+\overbar{\P}^{\rm max}_{\rm LTE,det}(\alpha_{n,\rm max}^{\rm det})=\P_{\rm driving}(\alpha_{n,\rm max}^{\rm det})$ gives
\begin{align}
&\frac{b(\alpha_{n,\rm max}^{\rm det}+1)^4}{\left[2\alpha_{n,\rm max}^{\rm det}+1-\sqrt{\alpha_{n,\rm max}^{\rm det}(3\alpha_{n,\rm max}^{\rm det}+2)}\right]^2} + 3\alpha_{n,\rm max}^{\rm det} \notag\\
&\qquad\qquad\qquad\qquad =  1+ 2\frac{3(\alpha_{n,\rm max}^{\rm det})^2+\alpha_{n,\rm max}^{\rm det}+(1-\alpha_{n,\rm max}^{\rm det})\sqrt{3(\alpha_{n,\rm max}^{\rm det})^2+2\alpha_{n,\rm max}^{\rm det}}}{2\alpha_{n,\rm max}^{\rm det}+1-\sqrt{3(\alpha_{n,\rm max}^{\rm det})^2+2\alpha_{n,\rm max}^{\rm det}}}.
\label{eq:BMcritLTE}  
\end{align}
For $b \to 1$, this can be approximated by~\cite{Ai:2023see} 
\bea 
\alpha^{\rm det}_{n,\rm max} \approx \frac{1-b}{3} \bigg(1+ \frac{4}{3}\sqrt{\frac{1-b}{6}}\bigg)\quad {\rm for\ } b\rightarrow 1.
\label{eq:BMcritLTEapp}
\eea 
In Fig.~\ref{fig:alphamax_det_munu} we show the comparison between the solution of Eq.~\eqref{eq:BMcritLTE} (blue solid line) and its approximation~\eqref{eq:BMcritLTEapp} (blue dashed line).

\subsection{\texorpdfstring{$\mu\nu$}{TEXT}-model}

The analysis for the bag EoS can be generalized to the $\mu\nu$-model~\cite{Leitao:2014pda}, which is also called the $\nu$-model in Ref.~\cite{Giese:2020rtr} and the template model in Refs.~\cite{Giese:2020znk,Ai:2023see,Sanchez-Garitaonandia:2023zqz}. The $\mu\nu$-model is a generalization of the bag model where the sound speed is allowed to deviate from $1/\sqrt{3}$. Explicitly, the EoS reads 
\begin{align}
\label{eq:nu_eos}
    &e_s(T)=\frac{1}{3} a_+ (\mu-1) T^\mu+\epsilon\,,\qquad p_s(T)=\frac{1}{3}a_+ T^\mu-\epsilon\,,\\
    &e_b(T)=\frac{1}{3}a_- (\nu-1)T^\nu \,,\quad\qquad\ \  p_b(T)=\frac{1}{3}a_- T^\nu\,,
\end{align}
where the constants $\mu$, $\nu$ are related to the sound speed in the symmetric and broken phases 
through
\begin{align}
    \mu=1+\frac{1}{c^2_{s}}\,,\quad \nu=1+\frac{1}{c^2_{b}}\,.
\end{align}

In this case the two matching conditions~\eqref{eq:conditionA} and~\eqref{eq:conditionB2} have a more specific form~\cite{Ai:2023see}
\begin{subequations}
\begin{align}
    \label{eq:mathching1}
    \frac{v_+}{v_-} &=\frac{v_+v_-(\nu-1)-1+3\alpha_+  }{v_+v_-(\nu-1)-1 +3v_+v_-\alpha_+},\\
    \label{eq:matching2}
    v_+v_- & = \frac{-\left(\frac{\gamma_+}{\gamma_-}\right)^{\nu} \Psi_+ + 1 - 3 \alpha_{+} }{3 \nu \alpha_{+}} \left[1-(\nu-1) v_+ v_-\right],
\end{align}
\end{subequations}
where 
\begin{align}
    \alpha_+=\alpha(T_+)\equiv \frac{\frac{1}{3} a_+(\mu-\nu) T_+^\mu+\nu \epsilon }{a_+ \mu  T_+^\mu}.
\end{align}
and $\Psi$ is the ratio of the enthalpies in the broken and the symmetric phase
\bea 
\Psi(T)\equiv\frac{\omega_b(T)}{\omega_s(T)}
\eea 
such that
\begin{align}
    \Psi_+\equiv \Psi(T_+)=\left(\frac{a_-}{a_+}\right)\left(\frac{\nu}{\mu}\right) T_+^{\nu-\mu}.
\end{align}
For $\mu=\nu=4$, $\Psi_+$ becomes the parameter $b$, and $\alpha_+$ becomes $\epsilon/(a_+ T_+^4)$.

In the detonation regime for the $\mu\nu$-model, we have $\Psi_+=\Psi_n\equiv (a_-/a_+)(\nu/\mu) T_n^{\nu-\mu}$. 
The Jouguet velocity is obtained by substituting $v_-=c_b$ into Eq.~\eqref{eq:mathching1} and reads~\cite{Giese:2020rtr,Ai:2023see}
\begin{align}
\label{eq:Jouguet-velocity-munu}
    \xi_J=c_{b}\left(\frac{1+\sqrt{3\alpha_{n}(1-c^2_{b}+3c^2_{b} \alpha_{n})}}{1+3c^2_{b}\alpha_{n}}\right)\,.
\end{align}
For $c_b=1/\sqrt{3}$, one recovers Eq.~\eqref{eq:Jouguet-velocity-bag}.  Substituting the EoS into Eq.~\eqref{eq:LTE-pressure2} and using $v_-=c_b$, $v_+=\xi_J$, one obtains 
\begin{align}
    \overbar{\P}_{\rm LTE, det}^{\rm max}=a_+T_n^{\mu}\left(\frac{\mu}{3}[\gamma(\xi_J)]^2 \xi_J(\xi_J-c_b)\right).
\end{align}
At the Jouguet point, $-\Delta V_T$  reads
 \begin{align}
    (-\Delta V_T)|_{\xi_J}=\frac{a_+ T_n^{\mu}}{3}\left[1- \left(\frac{\mu}{\nu} \right)\left(\frac{\gamma(\xi_J)}{\gamma(c_b)}\right)^\nu  \Psi_n\right],
\end{align}
where we also used the LTE matching condition~\eqref{eq:LTE-matching}. The driving pressure reads
\begin{align}
    \P_{\rm driving}=a_+ T_n^{\mu}\left[\frac{1}{3} +\left(\frac{\mu}{\nu}\right)\left(\alpha_n-\frac{1}{3}\right) \right].
\end{align}

Equaling $\P^{\rm max}_{\rm LTE,det}$ to $\P_{\rm driving}$, one can obtain $\alpha_{n,\rm max}^{\rm det}$. For $\Psi_n\rightarrow 1$, one has the approximation~\cite{Ai:2023see}

\begin{align}
\label{eq:alphamax_app_munu}
    \alpha_{n,\max}^{\rm det}\approx \frac{1-\Psi_n}{3}\left(1+ \frac{\nu}{3}\sqrt{\frac{1-\Psi_n}{(\nu-1)(\nu-2)}} \right) \quad {\rm for\ } \Psi_n \rightarrow 1 .
\end{align}

\begin{figure}[ht]
    \centering
    \includegraphics[scale=0.6]{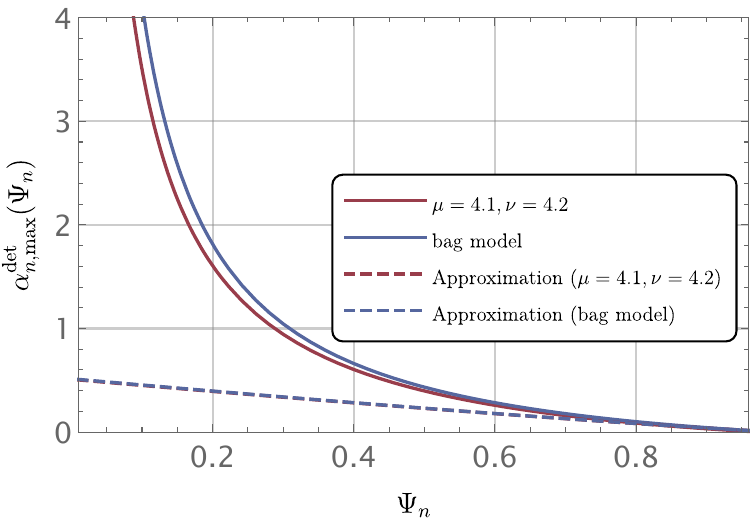}
    \caption{$\alpha_{n,\rm max}^{\rm det}$ as a function of $\Psi_n$ for the bag model ($\mu=\nu=4$, blue line) and for $\mu=~4.1,\nu=4.2$, and their approximations given in Eq.~\eqref{eq:alphamax_app_munu}. The lines for the approximations almost overlap for the chosen value of $\nu$. }
    \label{fig:alphamax_det_munu}
\end{figure}

For $\mu=4.1$, $\nu=4.2$, we show the comparison between the exact result (red solid) and the approximation (red dashed) in Fig.~\ref{fig:alphamax_det_munu}. In the numerical result, we also observe that, different from the approximation~\eqref{eq:alphamax_app_munu}, $\P^{\rm max}_{\rm LTE,det}=\P_{\rm driving}$ can have a {\it detonation} solution for $\alpha_{n,\rm max}^{\rm det}>0$ only when $\Psi_n$ is smaller than a critical value $\Psi_{n,\rm crit}$. For $\mu=4.1, \nu=4.2$, $\Psi_{n,\rm crit}\approx 0.97619$. For $\mu=\nu=4$, this critical value goes to one, as expected.

\section{Maximal hydrodynamic obstruction}
\label{sec:maximal_hydrodynamic_obstruction}

Now we generalize the analysis given in the last section to the hybrid regime and derive the maximal phase transition strength (for a given value of $b$ or $\Psi_n$) that allows for non-runaway hybrid motions, $\alpha_{n,\rm max}^{\rm hyb}$. According to the discussion in Sec.~\ref{sec:match-conditions}, this also gives the critical phase transition strength above which the maximal LTE force can never balance out the driving force, i.e., $\alpha_{n,\rm crit}^{\rm LTE}=\alpha_{n,\rm max}^{\rm hyb}$. 

\begin{figure}
    \centering
    \includegraphics[scale=0.3]{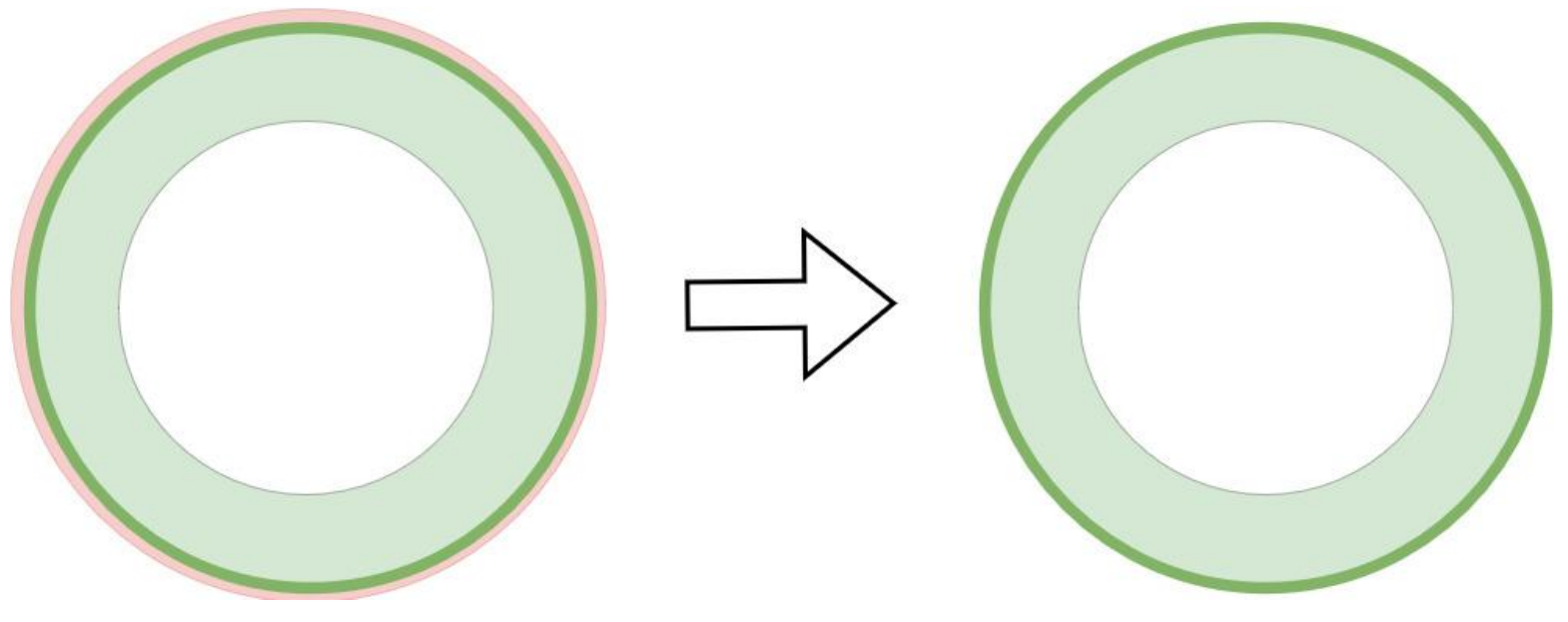}
    \caption{Transition from a hybrid solution to a detonation solution at the Jouguet point $\xi_w\rightarrow \xi_J$. As $\xi_w$ approaches $\xi_J$, the shock wave (light red region) becomes thinner and thinner until it disappears when $\xi_w=\xi_J$. The light green band represents the rarefaction wave behind the wall.}
    \label{fig:Jouguet_Transition}
\end{figure}

Below, we shall work with the $\mu\nu$-model directly and the results for the bag model are obtained by taking $\mu=\nu=4$. With this EoS and in the hybrid regime, we have 
\begin{subequations}
\label{Eq:Initial}
\begin{align}
\label{eq:Initial-1}
    \overbar{\P}_{\rm LTE}^{\rm max}&=a_+T_+^{\mu}\left[\frac{\mu}{3}[\gamma(v_+)]^2 v_+ (v_+ - c_b)\right],\\
    \left.(-\Delta V_T)\right|_{\xi_w=\xi_J} & = \frac{a_+ T_+^{\mu}}{3}\left[1- \left(\frac{\mu}{\nu}\right) \left(\frac{\gamma(v_+)}{\gamma(c_b)}\right)^\nu \Psi_+ \right],\\
    \P_{\rm driving} &= a_+ T_+^{\mu}\left[\frac{1}{3} +\left(\frac{\mu}{\nu}\right)\left(\alpha_+-\frac{1}{3}\right) \right].
\end{align}     
\end{subequations}
The common factor $a_+ T_+^\mu$ will be canceled out in the equation $\P^{\rm max}_{\rm LTE}=\P_{\rm driving}$.
Compared with the detonation regime, in the hybrid regime, we now have an additional shock wave. As a consequence, we do {\it not} have the good properties $v_+=\xi_w$ and $T_+=T_n$ ($\Rightarrow \alpha_+=\alpha_n$, $\Psi_+=\Psi_n$) anymore. One then needs to find the corresponding modified relations with the help of the matching conditions across the shock-wave front. Since $\P_{\rm LTE}^{\rm max}$ occurs nearly at $\xi_J$,
\begin{align}
\label{eq:xiw-hyb}
    \xi_w \approx \xi_J=c_b \left(\frac{1+\sqrt{3\alpha_n\left(1-c_b^2+3c_b^2 \alpha_n\right)}}{1+3 c_b^2\alpha_n}\right),
\end{align}
$\alpha_{n,\rm max}^{\rm hyb}/\alpha_{n,\rm crit}^{\rm LTE}$ can be derived by considering an infinitely thin shock wave. For larger $\xi_w$, the shock wave disappears and the motion enters into the detonation regime. This transition is schematically shown in Fig.~\ref{fig:Jouguet_Transition}. Because of the sudden appearance/disappearance of the shock wave when $\xi_w$ crosses $\xi_J$, $v_+$ and $\alpha_+$ are discontinuous at the transition. For more details, see Ref.~\cite{Espinosa:2010hh}, in particular Figure~6 there.

We denote the hydrodynamic quantities on both sides of the shock-wave front {\it in the rest frame of the latter} as $T_{\rm sw,\pm}$, $v_{\rm sw, \pm}$. To understand the physics of the hydrodynamic obstruction, we first realize that the shock wave reheats the plasma to some temperature $T_{\rm sw,-} > T_n$. In principle, the fluid equations~\cite{Espinosa:2010hh} should be integrated from $T_{\rm sw,-}$ to obtain the temperature right in front of the wall $T_+$. However, as the velocity of the wall approaches the Jouguet velocity $\xi_w \to \xi_J $, the distance between the shock-wave front and the wall diminishes to zero and $T_{\rm sw,-} \to T_+$. So, to study the pressure at the peak, we can set $T_{\rm sw,-} = T_+$. We will see that $T_+$ becomes fixed as a function of $\alpha_n$. Similarly, the fluid velocity right behind the shock-wave front $v_{\rm sw,-}$, and that right in front of the wall $v_{+}$, when transformed to the common plasma frame, are also equal to each other for $\xi_w\rightarrow \xi_J$. Therefore, we have 
\begin{align}
\label{eq:relations-hybrid}
    T_{\rm sw,+}=T_n,\qquad v_{\rm sw,+}=\xi_{\rm sw},\qquad T_{\rm sw,-}=T_+,\qquad \overbar{v}_{\rm sw,-}=\overbar{v}_+,
\end{align}
where $\xi_{\rm sw}$ is the velocity of the shock-wave front, and a bar indicates that the quantity is transformed to the plasma frame:\footnote{Note that all the velocity quantities are defined in a way such that they are positive. This needs to be taken into account when taking the Lorentz transformation. } 
\begin{align}
\label{eq:overbar-v}
    \overbar{v}_{\rm sw,-}\equiv\frac{\xi_{\rm sw}-v_{\rm sw,-}}{1-\xi_{\rm sw} v_{\rm sw,-}},\qquad \overbar{v}_+ \equiv\frac{\xi_w-v_+}{1-\xi_w v_+}.
\end{align}
The last relation in Eq.~\eqref{eq:relations-hybrid} gives~\cite{Ai:2021kak}
\begin{align}
\label{eq:tildev-2}
 v_{\rm sw, -}=\frac{v_+-\xi_w+\xi_{\rm sw}-v_+\xi_{\rm sw}\xi_w}{1+v_+ (\xi_{\rm sw}-\xi_w)-\xi_w \xi_{\rm sw}}, \quad \xi_{\rm sw}>\xi_w.
\end{align}

With the $\mu\nu$-model,  we have the following matching conditions at the shock-wave front~\cite{Ai:2023see}
\begin{subequations}
\begin{align}
&\xi_{\rm sw}=c_s\sqrt{\frac{(\mu-1)+\tilde{r}}{1+(\mu-1)\tilde{r}}}, \label{eq:sw-matching1}\\
&v_{\rm sw,-}=\frac{c_s^2}{\xi_{\rm sw}}\,,\label{eq:sw-matching2}
\end{align}
\end{subequations}
where  
\begin{align}
\label{eq:heating}
\tilde{r} \equiv \left(\frac{T_n}{T_+}\right)^\mu, \qquad \text{(heating parameter)} . 
\end{align} 
Looking at Eq.\eqref{eq:sw-matching1}, we observe that the maximal heating (smallest $\tilde{r}$) is obtained for the largest $\xi_{\rm sw}$ reachable.
Furthermore, one can show that
\begin{subequations}
\begin{align}
\label{eq:alpha+-alphan}
\alpha_+ &= \frac{\mu-\nu}{3\mu}+\left( \alpha_n -\frac{\mu-\nu}{3\mu} \right)\Tilde{r}\,,  \\
\label{eq:Psi+Psin}
\Psi_+ &= \Psi_n \tilde{r}^{1-\nu/\mu}. 
\end{align}
\end{subequations}
Combining Eqs.~\eqref{eq:tildev-2} and~\eqref{eq:sw-matching2} gives 
\begin{align}
\label{eq:xisw-v+-xiw}
    \frac{c_s^2}{\xi_{\rm sw}}=\frac{v_+-\xi_w+\xi_{\rm sw}-v_+\xi_{\rm sw}\xi_w}{1+v_+ (\xi_{\rm sw}-\xi_w)-\xi_w \xi_{\rm sw}}.
\end{align}
This relation can be used to show that the velocity of the shock-wave front is an increasing function of $\xi_w$, which means in turn that $\tilde r$ decreases (the heating increases) for a larger $\xi_w$, until the shock wave disappears at the Jouguet velocity. We thus confirm here that the peak in the pressure (due to heating) is located at the Jouguet velocity. 
Finally, $v_+$ can be eliminated by substituting $v_-=c_b$ into the matching conditions~\eqref{eq:conditionA} and~\eqref{eq:conditionB2} 
giving\footnote{Note the condition $v_+< v_-$ for hybrids so that one needs to select the correct root for the quadratic equation of $v_+$.}
\begin{align}
\label{eq:v+-alpha+}
    v_+=c_b \left(\frac{1-\sqrt{3\alpha_+\left(1-c_b^2+3c_b^2 \alpha_+\right)}}{1+3 c_b^2\alpha_+}\right).
\end{align}

Substituting Eqs.~\eqref{eq:xiw-hyb},~\eqref{eq:sw-matching1},~\eqref{eq:alpha+-alphan} and~\eqref{eq:v+-alpha+} into Eq.~\eqref{eq:xisw-v+-xiw}, one gets an algebraic equation for $\tilde{r}$ and $\alpha_n$. After obtaining $\tilde{r}(\alpha_n)$, one also numerically obtains $\{v_+,\xi_w,\xi_{\rm sw},\alpha_+\}$ in terms of $\alpha_n$. Substituting them into Eq.~\eqref{eq:Initial-1} and rewriting $\alpha_n$ as $\alpha_{n,\rm crit}^{\rm LTE}$, we then obtain $\overbar{\P}_{\rm LTE}^{\rm max}(\alpha_{n,\rm crit}^{\rm LTE})/(a_+ T_+^\mu)$. Together with Eq.~\eqref{eq:Psi+Psin}, we can also obtain $(-\Delta V_T)/(a_+ T_+^\mu)$, $\P_{\rm driving}/(a_+ T_+^\mu)$ in terms of $\alpha_n$. Finally, the equation $\P^{\rm max}_{\rm LTE}(\alpha_{n,\rm crit}^{\rm LTE})=\P_{\rm driving}(\alpha_{n,\rm crit}^{\rm LTE})$ gives $\alpha_{n,\rm crit}^{\rm LTE}$ as a function of $\Psi_n$.

\begin{figure}[ht]
    \centering
    \includegraphics[scale=0.6]
    {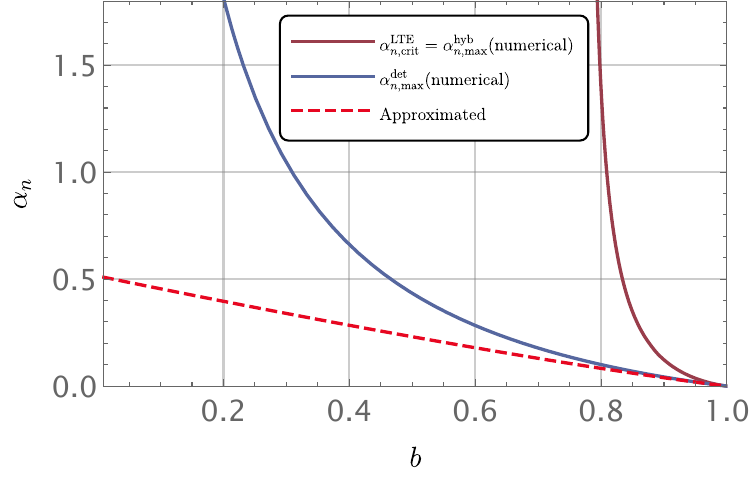}
    \includegraphics[scale=0.6]{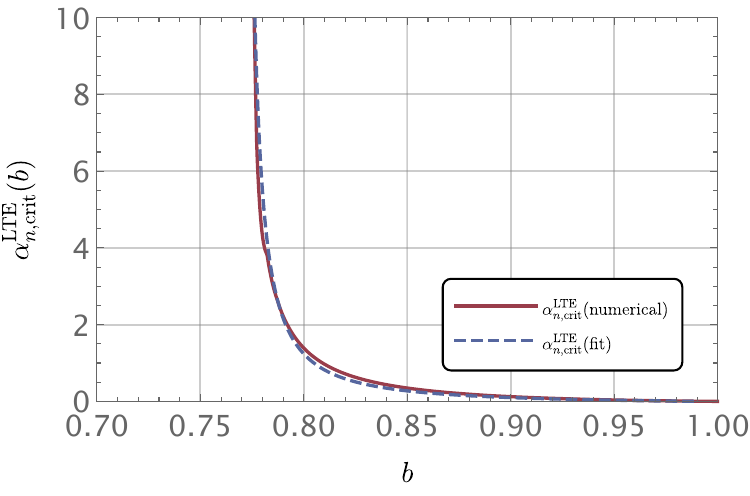}
    \caption{Left: Comparison between $\alpha_{n,\rm crit}^{\rm LTE}$, $\alpha_{n,\rm max}^{\rm det}$ and the approximation~\eqref{eq:BMcritLTEapp} for $\mu=\nu=4$. Right: Comparison between the numerical $\alpha_{n,\rm crit}^{\rm LTE}$ and the fit given in Eq.~\eqref{eq:alpha_crit_fit}.  }
    \label{fig:alpha_max_hyb}
\end{figure}

\paragraph{Bag equation of state.} We apply this numerical procedure for $\mu=\nu=4$ ($\Psi_n=b$ in this case) and show the result in Fig.~\ref{fig:alpha_max_hyb} (left panel) where we also compare the obtained $\alpha_{n,\rm crit}^{\rm LTE}$ with $\alpha_{n,\rm max}^{\rm det}$ and its approximation~\eqref{eq:BMcritLTEapp}. One can see that $\alpha_{n,\rm crit}^{\rm LTE}$ increases very fast as $b$ decreases. Unless when $b$ is very close to one, $\alpha_{n,\rm crit}^{\rm LTE}$ significantly differs from $\alpha_{n,\rm max}^{\rm det}$. We emphasize that $\alpha_{n,\rm crit}^{\rm LTE}(b)$ as a function of $b$ is universal; it does not depend on the underlying particle physics model. Therefore, for the convenience of future applications, we provide the following fit, valid for $\mu = \nu =4$,
\bea 
\label{eq:alpha_crit_fit}
\alpha_{n,\rm crit}^{\rm LTE}(b) \approx A + B (b - D)^{C}, \quad A = -0.1, \quad B =0.014, \quad C = -1.3, \quad D = 0.77.
\eea
The comparison between the above fit and the numerical result is shown in the right panel of Fig.~\ref{fig:alpha_max_hyb}.

\begin{figure}[ht]
    \centering
    \includegraphics[scale=0.8]{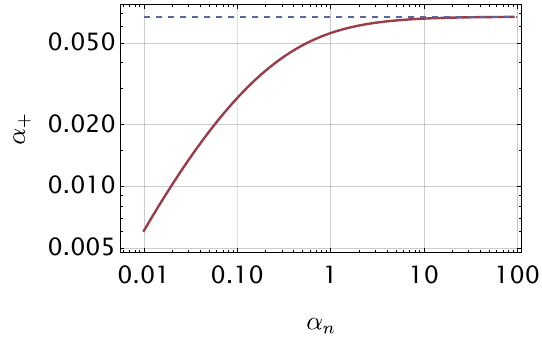}
    \includegraphics[scale=0.8]{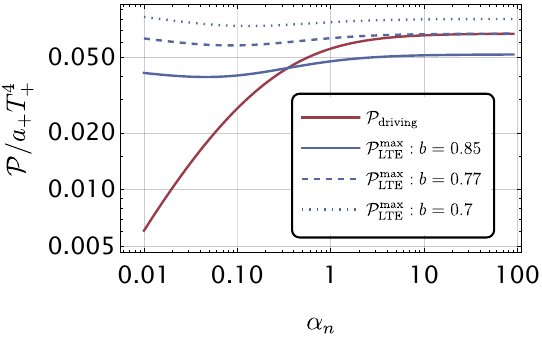}
    \caption{Left: Plot of $\alpha_+$ as a function of $\alpha_n$ in the limit $\xi_w\rightarrow\xi_J$ from the hybrid branch. We observe a saturation at  $\alpha_+(\alpha_n \to \infty) \to 0.0668$. This saturation indicates that from the point of view of the phase boundary, the effective phase transition strength saturates as we increase $\alpha_n$. Right: Comparison between the driving and maximal LTE pressure on the bubble wall: When the maximal LTE friction (blue line) is always higher than the driving force (Red line), the heating prevents the runaway, for any $\alpha_n$. }
    \label{fig:alphap_vs_alphan}
\end{figure}

Moreover, it seems that there is a critical value of $b$, around $0.77$, where $\alpha_{n,\rm crit}^{\rm LTE}$ diverges. To clarify the physical significance of this divergence, we plot on the Left of Fig.~\ref{fig:alphap_vs_alphan} the value of the $\alpha_+$, as a function of $\alpha_n$, in the limit of $\xi_w\rightarrow \xi_J$ from the hybrid branch. This shows that \emph{within the LTE approximation}, no matter how supercooled the phase transition is, in the limit of large $\alpha_n$, the effective strength that the phase boundary feels goes to a constant $\alpha_+(\alpha_n \to \infty) \to 0.0668$. This means that using the equations for the driving pressure in Eq.~\eqref{Eq:Initial}, the driving force saturates, even for $\alpha_n$ going to infinity. On the Right of Fig.~\ref{fig:alphap_vs_alphan}, we show that the driving force never reaches the friction for $b<0.77$. It is expected that the upper bound on $\alpha_+$ at the (hybrid) Jouguet point should be uplifted when the LTE condition is relaxed. However, it is well known that $\alpha_+<1/3$ for any deflagration/hybrid solutions regardless of whether the LTE condition is satisfied at the wall or not. 

This result agrees with the numerical analysis presented in Ref.~\cite{Ai:2023see}. Since the peak in the pressure essentially comes from a heating effect in front of the bubble wall, which is absent in the limit of $\alpha_n \to \infty$, this result must be unphysical. Our analysis may thus break down at some large value of $\alpha_n$ and the unphysical conclusion should be mitigated when the hydrodynamics analysis is replaced with an analysis in terms of microphysical particle processes with an exchange of momentum between the wall and the plasma. How exactly the analysis may break down as $\alpha_n$ increases is left for future work, but will be very briefly discussed in Sec.~\ref{sec:comment_on_LTE}.

\begin{figure}[ht]
    \centering
    \includegraphics[scale=0.8]{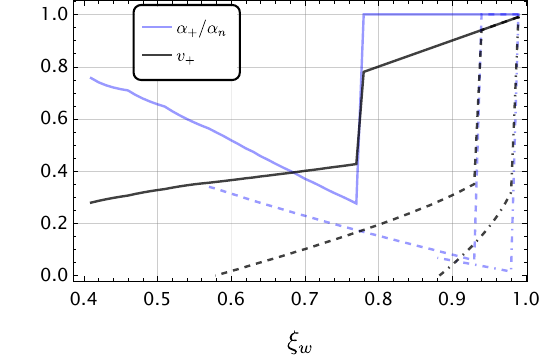}
    \includegraphics[scale=0.8]{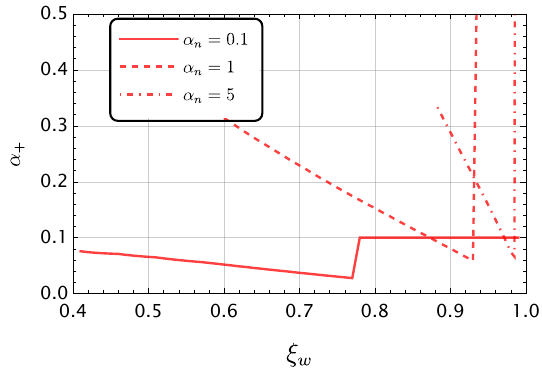}
    \caption{Left: Evolution of $v_+$ (black) and $\alpha_{+}/\alpha_n$ (blue) for $\alpha_n = 0.1$ (solid lines), $\alpha_n = 1$ dashed lines) and $\alpha_n = 5$ (dot-dashed lines).  Right: Evolution of $\alpha_+(\xi_w)$ for for $\alpha_n = 0.1$ (solid lines), $\alpha_n = 1$ dashed lines) and $\alpha_n = 5$ (dot-dashed lines). We set $c_b = c_s = 1/\sqrt{3}$. %The gridline represent different values of $\alpha_{+}$.  
    }
    \label{fig:solving_self_similar}
\end{figure}

\paragraph{Going away from the Jouguet point.}
Previously we have observed that the hydrodynamic obstruction occurs mostly due to a dramatic heating close to the Jouguet point. 
To justify our picture and to see how the heating grows when $\xi_w$ approaches $\xi_J$ from below, we also solve the fluid equations and search for the self-similar fluid temperature and velocity profiles for any $\xi_w$ approaching $\xi_J$. On the left panel of Fig.~\ref{fig:solving_self_similar}, we present the evolution of $v_+$ (black) and $\alpha_{+}/\alpha_n$ (blue) for $\alpha_n = 0.1$ (solid lines), $\alpha_n = 1$ (dashed lines) and $\alpha_n = 5$ (dot-dashed lines). We observe that the maximal heating occurs indeed at the Jouguet velocity, where both $\alpha_+(\xi_w)$ and $v_+(\xi_w)$ are discontinuous. On the right panel, we show the evolution of $\alpha_+(\xi_w)$. Its values on the hybrid side of the discontinuity confirm the saturation predicted and observed on the left panel of Fig.~\ref{fig:alphap_vs_alphan}.

\section{Comparison with the B\"{o}deker-Moore criterion}
\label{sec:compare_with_BM}

The BM thermal frictional pressure is given in Eq.~\eqref{eq:BM-force}. This pressure is model dependent. Defining a characteristic energy scale $M$
\begin{align}
    M^2 \equiv \frac{\sum_i c_i g_i \Delta m^2_i}{\sum_i g_i}\,,
\end{align}
where the sum goes over the species that acquire a mass through the phase transition,
Eq.~\eqref{eq:BM-force} becomes 
\begin{align}
    \P_{\rm BM}\approx \frac{\Delta g_{\rm eff} M^2 T_n^2  }{24},
\end{align}
where $\Delta g_{\rm eff}$ denotes the change in the number of relativistic DoFs in the plasma from the outside to the inside of the wall. We have assumed that the particles gaining mass have $m_i \gg T_-$ inside the wall such that they do not contribute to the relativistic DoFs after entering the bubble. To compare it with $\P_{\rm LTE}^{\rm max}$, we consider the bag EoS where $(a_+ - a_-) =\pi^2 \Delta g_{\rm eff}/30$ and we have 
\begin{align}
   \P_{\rm BM} \approx \frac{30 M^2 T_n^2  (a_+-a_-) }{24\pi^2}= \frac{5M^2 T_n^2 (1-b) a_+}{4\pi^2}. 
\end{align}
Therefore, we have the BM runaway criterion
\begin{align}
    \frac{5(1-b)}{4 \pi^2} \left(\frac{M}{ T_n}\right)^2 (a_+ T_n^4) < \alpha_n (a_+ T_n^4) \equiv \Delta V \quad \Rightarrow\quad \alpha_n> \frac{5(1-b)}{4\pi^2} \left(\frac{M}{ T_n}\right)^2\equiv \alpha_{n,\rm crit}^{\rm BM},
\end{align}
where we have used the expression of the driving pressure in the bag EoS. The parameter $M/T_n$ can be calculated once the particle physics model is given.

\begin{figure}[ht]
    \centering
    \includegraphics[scale=0.6]{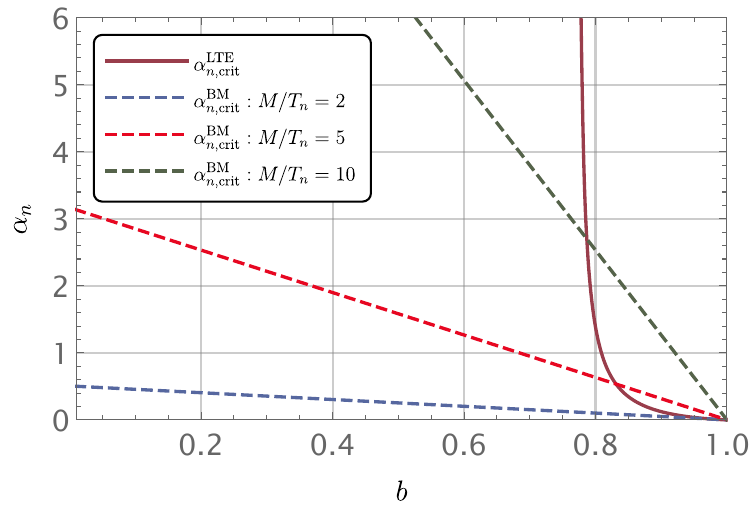}
    \includegraphics[scale=0.6]{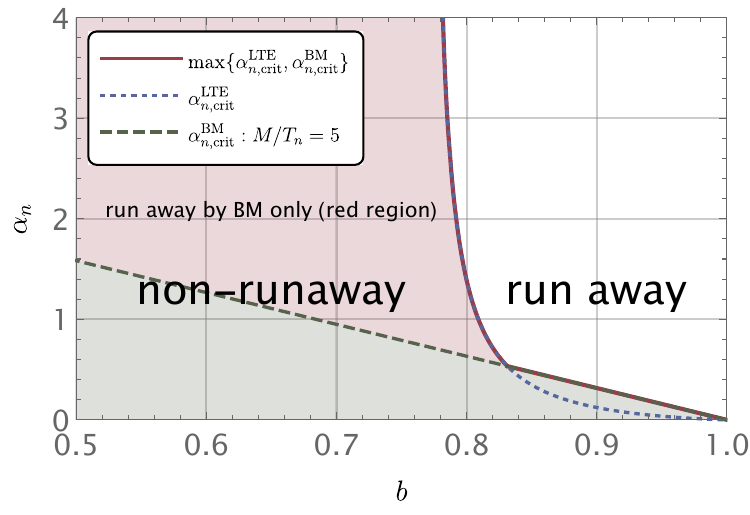}
    \caption{Left: Comparison between $\alpha_{n,\rm crit}^{\rm LTE}$ and $\alpha_{n,\rm crit}^{\rm BM}$ for $\mu=\nu=4$ and $M/T_n=2, 5, 10$. Right: Modified runaway curve obtained from ${\rm max}\{\alpha_{n,\rm crit}^{\rm LTE},\alpha_{n,\rm crit}^{\rm BM}\}$, above which the bubble wall's acceleration can be stopped by neither the hydrodynamic obstruction nor the leading-order BM thermal friction. All the shaded regions (red and green) give non-runaway walls by the modified criterion, while the red region would give runaway walls if hydrodynamic obstruction is not taken into account. Recall our comment below Eq.~\eqref{eq:BMcriterion} on the term ``runaway'' used in this paper.}
    \label{fig:LTE_vs_BM}
\end{figure}

In the left panel of Fig.~\ref{fig:LTE_vs_BM}, we compare $\alpha_{n,\rm crit}^{\rm LTE}$ with $\alpha_{n,\rm crit}^{\rm BM}$ for $M/T_n=2, 5, 10$. In the right panel of Fig.~\ref{fig:LTE_vs_BM}, taking $M/T_n=5$ as an example, we show the modified runaway criterion. The dashed and dotted lines represent the BM and LTE runaway criteria, respectively. The solid line, which is obtained from ${\rm max}\{\alpha_{n,\rm crit}^{\rm LTE},\alpha_{n,\rm crit}^{\rm BM}\}$, represents the modified runaway criterion. The shaded region gives a conservative estimate for the parameter space allowing non-runaway walls. The large red region would be (incorrectly) estimated to give runaway walls if hydrodynamic obstruction is not taken into account, even in the non-supercooled regime $\alpha_n < 1$. 

Note that for $\alpha_n< \alpha_{n,\rm crit}^{\rm LTE}$, the wall velocity is bounded by the Jouguet velocity, $\xi_w\leq \xi_J$. For a phenomenologically large $\alpha_n$, e.g., $\alpha_n=10$, $\gamma(\xi_J)\approx 7.64$, which is much smaller than the typical value of $\gamma_w$ for the ultrarelativistic regime considered in Refs.~\cite{Bodeker:2009qy,Bodeker:2017cim}, $\gamma_w\sim 10^9$. Actually, the phase transition strength cannot be arbitrarily large, otherwise, the phase transition cannot complete~\cite{Ellis:2018mja}. This ensures that the LTE force becomes maximal typically before the wall enters the ultrarelativistic regime for practical phase transition strengths. Therefore, $\alpha_{n,\rm crit}^{\rm LTE}$ alone can serve as a criterion for whether the wall can enter into the ultrarelativistic regime or not, and thus is important for phenomenological studies involving ultrarelativistic walls. Let us now illustrate within an actual particle physics model, how to compute the hydrodynamic obstruction.

\paragraph{How to compute $b$ in BSM models.}

We have seen that $b\equiv a_-/a_+$ is a crucial quantity for the determination of the hydrodynamic obstruction. Let us now discuss how to work out its value for BSM particle physics models, especially if the BSM sector is weakly coupled with the SM (as considered in, e.g., Ref.~\cite{Fairbairn:2019xog}). First of all, the matching conditions are based on the total energy-momentum conservation between the order-parameter scalar and the plasma. If some particle species do not respond to the passage of the wall quickly enough, they can be taken as completely ``invisible'' to the wall and they should not be taken into account when analyzing the total energy-momentum conservation.
Therefore, $a_+$ should contain all the particles that couple directly/strongly to the bubble wall (that we call \emph{active}) which give a contribution $a_{+, \rm active}$. However, there are some particles that couple {\it indirectly} to the wall via the active particles (that we call \emph{passive}) such that the exchange of energy/momentum between the active and the passive particles occurs on a length scale much smaller than the size of the perturbed fluid around the wall, which can be approximated by the bubble radius.\footnote{MV thanks Benoit Laurent for very helpful discussions on this point.} These passive particles also need to be counted and give a contribution $a_{+,\rm passive}$. Mathematically, we have 
\begin{align}
a_{+, \rm passive} &= \text{number of DoF with } R_{\rm bubble} \gg L_{\rm therm},\notag 
\\ 
a_+ &= a_{+, \rm active}+ a_{+, \rm passive},\notag
\end{align}
where $L_{\rm therm}$ is the length of thermalisation of the passive particles due to the exchange of energy/momentum between active and passive particles. In the opposite limit, particles coupling very weakly to active ones such that $R_{\rm bubble} \ll L_{\rm therm}$ can be fully ignored in the analysis. As an example, in the EWPT, the top quark would be an active particle, the electron would be a passive particle\footnote{In this example, the electron does interact with the Higgs directly. But this direct Yukawa coupling is very small and hence can be ignored. However, the electron can interact with Higgs efficiently via the weak interactions.} and the neutrino would be discarded in counting the DoFs. The quantity $a_-$ is computed in the same way, with the further requirement that $m_- < T_-$. Since the passive particles do not couple to the order parameter, their DoFs do not change across the wall as long as the change in the temperature across the wall is not too big:
\bea 
a_{+, \rm passive} = a_{-, \rm passive}. 
\eea

\paragraph{Relevance for particle physics models and phenomenology.}

Let us quickly comment on a typical model where our discussion might be relevant. Spontaneous $B-L$ symmetry breaking is a physically motivated phase transition with a Lagrangian of the following form
\begin{align}
\label{Eq:L_RHN}
\mathcal{L}_{\text{int}} = &\frac{1}{2} \sum_{I} 
%\frac{1}{\sqrt{2}}
y_I \Phi \overbar{N}_I^c N_I + \sum_{\alpha,\, I} Y_{D,\alpha I} H\overbar{L}_{\alpha}N_I +  h.c.,
\end{align}
where $L_\alpha$ are the SM lepton doublets, $N_I$ are the three 
families of heavy right-handed (RH) neutrinos, $Y_{D,\alpha I}$ are the \emph{Dirac} Yukawa couplings between $N_I$ and $L_\alpha$, and $y_I$ are \emph{Majorana} Yukawa couplings. After the phase transition, $\langle \Phi \rangle = v_\phi/\sqrt{2}$ and the type-I seesaw Lagrangian is recovered with $M_I = \frac{1}{\sqrt{2}}y_I v_\phi$.

The phase transition $\langle \Phi \rangle = 0 \to v_\phi/\sqrt{2}$ can be strongly first order if $\Phi$ couples to another scalar $s$ via a coupling of the form $-\Delta {\cal L} = \frac{\lambda_{s\phi}}{2} s^2 |\Phi|^2$, inducing a mass $M_s^2(\phi)=\lambda_{s\phi } \phi^2$. It has been shown that the bubble wall induces an enhancement of the leptogenesis efficiency~\cite{Chun:2023ezg}, which however strongly depends on the terminal velocity of the bubble wall. 
$b \equiv a_-/a_+$ can be computed in the following way: $ (30/\pi^2) a_{+, \rm active} = 6 + 2+1 = 9$ (assuming three families of RH neutrinos, one complex $\Phi$ and a singlet $s$),  and $(30/\pi^2) a_{-, \rm active} \approx 1$. $ a_{+, \rm passive}$ is more delicate to compute and depends on the value of $Y_{D,\alpha I}$. 
The $N_I$ couple to the SM via the couplings $Y_{D,\alpha I}$ and the typical length scale for thermalisation between $N_I$ (active particles) and SM particles, typically via Higgs-mediated $NL \to NL$, is given by~\cite{Kurkela:2011ti}
\bea 
L_{\rm therm}^{-1} \equiv \langle \sigma_{NL \to NL} v \rangle n_L \sim  Y_{D}^4\frac{T_n}{8\pi^3}.
\eea 
On the other hand, the largest bubble radius can be estimated as
\begin{align}
    R_{\rm max}\sim \frac{(8\pi)^{1/3} \xi_w}{\beta} \sim \frac{(8\pi)^{1/3}}{\beta}  .
\end{align}
where $\tilde \beta H \equiv \beta \equiv HT \frac{d}{dT}\big(S_3/T\big)$ is the timescale of the transition in Hubble units.
Within our model for $ M_I \sim T_n \sim 10^{10} $ GeV and $\tilde \beta \sim 10$, we have 
\bea 
Y_{D}^2 \sim 10^{-12} \frac{M_I}{10^3 \text{GeV}}\quad \Rightarrow\quad L_{\rm therm} \sim {10^{2}} \text{GeV}^{-1} \gg R_{\rm max} \sim 10^{-1} \text{GeV}^{-1}.
\eea 
So under perturbations, the SM particles never thermalise within a scale shorter than the bubble scale. Therefore, we have $a_{+, \rm passive} = a_{-, \rm passive} = 0$.  We obtain finally $b \sim 0.1$, which is a value that, according to the modified criterion, would not allow for runaway. In this case, we could approximate $\xi_w \approx \xi_J$.

\section{Comments on the validity of our approximations}
\label{sec:comment_on_LTE}

We have seen that the hydrodynamic obstruction is essentially a consequence of the strong heating of the plasma due to the shock wave in front of the wall. For this heating to occur, we however have to fulfil some conditions on hydrodynamics and exchange of energy.
In this section, we would like to discuss a few possible weaknesses of our results, for the purpose of further studies.  
\begin{itemize}
\item {\textbf{The hydrodynamic regime:}}
We would like to emphasize that in this paper, our computation relied on hydrodynamics, which is by definition valid for scales much larger than the mean free path of particles in the plasma. For example, for a gauge theory with coupling $g$, the mean free path of the particles coupling to the gauge field would be~\cite{Kurkela:2011ti} 
\bea
L_{\rm MFP} \equiv \frac{1}{\langle \sigma_{2 \to 2} v \rangle n_{\rm scatterers}} \propto \frac{4 \pi}{ g^4 T_{\rm n}} = \mathcal{O}(10) \times \bigg(\frac{\alpha_n}{\Delta V}\bigg)^{1/4} \frac{1}{g^4},
\eea 
which we evaluated at the nucleation temperature. 
As we have seen, in the LTE picture, the pressure as a function of the velocity is not a monotonic function, but features a peak in the pressure at $\xi_{\rm peak} \sim \xi_J$, which is related to $\alpha_n$ via Eq.~\eqref{eq:Jouguet-velocity-bag} for the bag EoS. This means that for $\alpha_n \gg 1$, 
$\gamma(\xi_{J}) \sim \mathcal{O}(1)\alpha_n^{1/2}$.

In order for the hydrodynamics analysis to be valid, one needs to require two conditions. First, the length scale of the perturbed fluid should be larger than the mean free path of the plasma. The former can be (very roughly) estimated as the bubble radius $R_w$ and hence we require
\begin{align}
\label{cond1}
   R_w \gg  L_{\rm MFP}.
\end{align}
This issue can be understood in another way, from the particle physics perspective. We have seen that the hydrodynamic obstruction disappears if the bubble wall manages to become a detonation. On the other hand, the hydrodynamic obstruction originates from the heating of the plasma due to the particles bouncing off the wall. For the hydrodynamic obstruction to be relevant in the early stage of the bubble wall, the time scale of the heating, $t_{\rm heating}$, needs to be smaller than the time scale for the bubble wall to accelerate to the Jouguet velocity, $t_{\rm acceleration}$:
\bea 
t_{\rm acceleration} \gg t_{\rm heating}.
\eea

Second, one may also require the thickness of the shock wave to be larger than $L_{\rm MFP}$ in order for the hydrodynamic description on the shock wave to be valid\footnote{We thank Filippo Sala and Aleksandr Azatov for very useful discussions about this condition},
\begin{align}
\label{cond2}
    R_{\rm sh}(\xi_w,\alpha_n)-R_w(\xi_w,\alpha_n) > L_{\rm MFP},
\end{align}
where $R_{\rm sh}$ is the radial position of the shock-wave front.
The equation $R_{\rm sh}(\xi_w^\star,\alpha_n)-R_w(\xi_w^\star,\alpha_n) = L_{\rm MFP}$ then defines a function $\xi_w^\star (\alpha_n)$. For $\xi_w^\star(\alpha_n)<\xi_w< \xi_J(\alpha_n)$, the thickness of the shock wave would be smaller than $L_{\rm MFP}$. Strong heating $\alpha_+/\alpha_n \ll 1$ require typically a short distance $R_{\rm sh}(\xi_w,\alpha_n)-R_w(\xi_w,\alpha_n)$, so we expect this condition to be more stringent than Eq.~\eqref{cond1}. Such a detailed analysis is beyond the scope of this paper and will be presented in another work~\cite{AiPrepa}.

\item The thin-wall approximation: In hydrodynamics, one has implicitly assumed a thin wall whose size can be neglected compared to the hydrodynamic scale. Therefore, the hydrodynamics analysis may be valid only when
\begin{align}
\label{cond3}
    \frac{L_w}{\gamma_w(\xi_w)} < R_{\rm sh}(\xi_w,\alpha_n)-R_w(\xi_w,\alpha_n).
\end{align}
The equation $ L_w/\gamma_w(\xi_w^\dagger) = R_{\rm sh}(\xi_w^\dagger,\alpha_n)-R_w(\xi_w^\dagger,\alpha_n)$ then defines another function $\xi_w^\dagger(\alpha_n)$. For $\xi_w^\dagger(\alpha_n)<\xi_w<\xi_J(\alpha_n)$, the thickness of the shock wave would be smaller than $L_{w}/\gamma_w$. 

It is then possible that one should evaluate the peak height at ${\rm min}\{\xi_w^\star(\alpha_n),\xi_w^\dagger(\alpha_n)\}$ instead of $\xi_J(\alpha_n)$. This should reduce the height of the pressure peak. A detailed analysis of all the three conditions~\eqref{cond1},~\eqref{cond2}~\eqref{cond3} and a study on the pressure when these conditions are violated go beyond this work and are left for future work.

\item {\textbf{The self-similar solution}}
Even in the regime of validity of hydrodynamics, there is no guarantee that the peak in the pressure will manifest. Although in our analysis, we do not need to solve the fluid equations, the classification of the fluid motions and the particular Jouguet point may still be based on \emph{self-similar solutions for the temperature and bulk velocity profiles}. It is known that such self-similarity profiles are reached asymptotically when $R_w \gg R_{\rm initial}$. However, in the early stage of the bubble wall, the temperature profile might differ substantially from the self-similar solution. This would occur if the time scales of adaptation of the profile is much longer than the timescale of acceleration of the bubble.\footnote{MV thanks Ryusuke Jinno for enlightening discussions on this point.}
Verifying the validity of the self-similar profile or the real ``existence'' of the pressure peak, requires dynamical simulation, a study that we will perform in a forthcoming study~\cite{AiPrepa}.  
\end{itemize}

\section{Conclusion}
\label{sec:conc}

In this paper, we have studied the criterion for the runaway behaviour of bubble walls within two opposite approximations: the ballistic approximation and the LTE approximation. These two approximations are usually applied to the ultrarelativistic regime and the relatively lower velocity regime, respectively.  The former corresponds to the BM thermal friction and the latter corresponds to the hydrodynamic obstruction. These frictions thus typically exist in different regions of the wall velocity space. For both approximations, we formulate a different runaway criterion in terms of a critical phase transition strength, $\alpha_{n,\rm crit}^{\rm BM/LTE}$. For $\alpha_n> \alpha^{\rm BM/LTE}_{n,\rm crit}$, the acceleration of the wall cannot be stopped by the friction under study. The BM criterion for runaway walls $\alpha_{n,\rm crit}^{\rm BM} <\alpha_n$, valid within the ballistic approximation, has been widely used in the literature. 

However, the BM criterion implicitly assumes that the friction is a monotonous function of the wall velocity. This is however not true and the hydrodynamic obstruction gives a pressure barrier before the wall enters into the ultrarelativistic regime~\cite{Cline:2021iff,Laurent:2022jrs,Ai:2023see}. 
The modified criterion on runaway when the hydrodynamic obstruction is taken into account should be $\alpha_n > {\rm max}\{\alpha_{n,\rm crit}^{\rm BM}, \alpha_{n,\rm crit}^{\rm LTE}\}$.
While both $\alpha_{n,\rm crit}^{\rm BM}$ and $\alpha_{n,\rm crit}^{\rm LTE}$ depend on the ratio of the number of DoFs in the symmetry and broken phases $b$, $\alpha_{n,\rm crit}^{\rm BM}$ further depends on a particle physics model-dependent parameter, the ratio between the averaged mass gain and the nucleation temperature $M/T_n$. Therefore, ${\rm max}\{\alpha_{n,\rm crit}^{\rm BM}, \alpha_{n,\rm crit}^{\rm LTE}\}$ depends on the interplay of these two parameters. We have compared $\alpha_{n,\rm crit}^{\rm BM}$ with $\alpha_{n,\rm crit}^{\rm LTE}$ for some typical values of the model-dependent parameter $M/T_n$ (cf. Fig.~\ref{fig:LTE_vs_BM}). We have argued that the BM is not always a sufficient criterion for the runaway (even in the absence of other sources of friction due to, e.g., $1 \to 2$ processes). In Sec.~\ref{sec:maximal_hydrodynamic_obstruction}, Eq.~\eqref{eq:alpha_crit_fit}, we provide easily implementable expressions for the maximal LTE pressure and we emphasize the fact that $\alpha_{\rm n,crit}^{\rm LTE}(b)$ \emph{does not} depend on the particularities of the underlying particle physics model. 
It appears that within the LTE approximation, theories with large enough change in DoF, $b < 0.77$, have typically non-runaway walls due to the hydrodynamic obstruction. This would reduce drastically the region of parameter space allowing for runaway. 
Furthermore, $\alpha_{\rm n,\rm crit}^{\rm LTE}$ is the lower bound for the phase transition strength that the wall can enter the ultrarelativistic regime and thus is a very important quantity for phenomenological studies involving ultrarelativistic walls.

We note that in this paper we have ignored other sources of friction that may exist. For the estimate of the pressure peak, we have ignored out-of-equilibrium effects which were shown to be not small in Refs.~\cite{DeCurtis:2022hlx,DeCurtis:2023hil,DeCurtis:2024hvh}, in contrast to Ref.~\cite{Laurent:2022jrs}. For the BM criterion, we have ignored other sources of friction due to other particle physics processes. As a consequence, both $\alpha_{n,\rm crit}^{\rm LTE}$ and $\alpha_{n,\rm crit}^{\rm BM}$ serve as a most conservative estimate of the maximal phase transition strength under which non-runaway stationary walls can exist.  

We also attract the attention to several caveats that may change our conclusions. Specifically, the velocity at which the hydrodynamic obstruction is maximal, $\xi_J$, is relatively low and can be reached at the early stage of the bubble expansion. Also, at the Jouguet velocity, the shock wave becomes infinitely thin. Whether the approximations or assumptions used in this paper, i.e., the LTE approximation, hydrodynamics, and the self-similarity used for the plasma temperature and velocity profiles, are valid or not in such an early stage, or when the shock wave becomes infinitely thin needs further investigation. We hope that our work can serve as a motivation to study the dynamical evolution of the bubble wall, not only its steady states, an investigation that we will initiate in Ref.~\cite{AiPrepa}.

\section*{Acknowledgments}

The authors would like to thank Aleksandr Azatov, Simone Blasi, Benoit Laurent, Jorinde Van de Vis and Carlos Tamarit for clarifying discussions, as well as Giulio Barni, Ryusuke Jinno and Filippo Sala for comments on the draft. The work of WYA is supported by the Engineering and Physical Sciences Research Council (grant No. EP/V002821/1). XN is supported by the iBOF “Un-locking the Dark Universe with Gravitational Wave Observations: from Quantum Optics to Quantum Gravity” of the Vlaamse Interuniversitaire Raad. MV is supported by the ``Excellence of Science - EOS" - be.h project n.30820817, and by the Strategic Research Program High-Energy Physics of the Vrije Universiteit Brussel.

\newpage

\appendix 

\section{Discussion on the departure from LTE}
\label{App:entropy_inj}

In the main text, we have always assumed LTE and neglected the increase of entropy at the phase boundary. In this appendix, we try to broaden this picture and discuss what the increase of entropy would change. 

\paragraph{Modification in the matching conditions.}
The first modification would appear in the relation between the pressure and the potential~\cite{Ai:2021kak} 
\bea
p(\phi,T) =  -V_{\rm eff}(\phi,T)+\delta p = - V(\phi)- V_T(\phi,T) + \delta p \equiv \Bar{p}(\phi,T)+\delta p,
\eea
where $\delta p$ encodes the out-of-equilibrium effects. We have introduced $\Bar{p} \equiv - V(\phi)- V_T(\phi,T) $ to denote the equilibrium part of the pressure. In this notation, the $p$ in the main text should be replaced with $\Bar{p}$. With this additional $\delta p$, the matching condition in Eq.~\eqref{eq:conditionB} is modified to (assuming stationary motion for simplicity for the present discussion)
\begin{align}
   \omega_+\gamma_+^2v_+^2+p_+=\omega_-\gamma_-^2v_-^2+p_-\Rightarrow \Delta V= \Delta(-V_T+\omega \gamma^2 v^2+\delta p).\label{eq:conditionB-modified}
\end{align}
Then one can identify the friction as
\begin{align}
    \P_{\rm friction}=\P_{\rm LTE}+\Delta \delta p,
\end{align}
where $\P_{\rm LTE}$ is defined as previously. Comparing it with Eq.~\eqref{eq:Pfriction_as_a_sum}, one then has
\begin{align}
    \Delta \delta p &=\P_{\rm dissipative}= - \int_{-\delta}^\delta \d z\, (\partial_z\phi)\left(\sum_i\frac{\d m^2_i(\phi)}{\d\phi}\int \frac{\d^3{\bf p}}{(2\pi)^32E_i}\,\delta f_i(p,z)\right),\notag\\
    &\Rightarrow  \frac{\d\delta p}{\d z}=- (\partial_z\phi)\left(\sum_i\frac{\d m^2_i(\phi)}{\d\phi}\int \frac{\d^3{\bf p}}{(2\pi)^32E_i}\,\delta f_i(p,z)\right).
\end{align}
The above equation can be used to compute $\delta p$ from $\delta f_i$, the out-of-equilibrium part of the distribution functions.

Allowing for out-of-equilibrium, thermodynamic identities may be spoiled as there is no well-defined temperature for $\delta f_i$. We will assume a close-to-equilibrium system such that $\delta f_i \equiv f - f_i^{\rm eq} \ll f_i^{\rm eq}$ so that we still use $f_i^{\rm eq}$ to define most thermodynamic quantities except the additional $\delta p$ in the pressure. In this sense, only the very direct effects on the friction of the wall are considered. Then, 
\begin{align}
    e(\phi,T)=-\Bar{p}(\phi,T)+T\frac{\partial \Bar{p}(\phi,T)}{\partial T},
\end{align}
as defined in LTE, and the entropy density is defined as $s=\omega/T=(e+\bar{p})/T$.

Although the entropy density is defined as previously, the presence of the deviation from equilibrium can induce the entropy production such that the matching condition Eq.~\eqref{eq:LTE-matching_1}  is modified by the introduction of a parameter $\Delta s$:
\bea 
(s_+ + \Delta s) \gamma_+ v_+ =s_- \gamma_- v_-.
\eea  
Using Eq.~\eqref{eq:conditionA}, it can also be written as
\begin{align}
 \label{eq:LTE-matching_bis}
\frac{T_+}{T_-}=\frac{\gamma_-}{\gamma_+}\bigg(1 + \frac{\Delta s}{s_+} \bigg)  \,.
\end{align}

It is clear that the parameters $\Delta s$ and $\delta p$ are in principle computable within a particle physics model and are related to each other.  However, the computation of those parameters from a specific particle physics model is beyond the scope of this work. We leave it for future work.

\paragraph{Consequence of the entropy increase.}
For $-\Delta V_T$, we have
\begin{align}
    -\Delta V_T= \frac{a_+T_n^4}{3} \left(1 -\frac{b r^4 }{(1+\Delta s/s_+)^{4}}\right),
\end{align}
and thus the different contribution of the pressure in Eq.~\eqref{Eq:Initial} becomes 
\begin{subequations}
\begin{align}
\label{Eq:Initial_bis}
    \overbar{\P}_{\rm LTE}^{\rm max}&=a_+T_+^{\mu}\left[\frac{\mu}{3}[\gamma(v_+)]^2 v_+ (v_+ - c_b)\right],\\
    \left.(-\Delta V_T)\right|_{\xi_w=\xi_J} & = \frac{a_+ T_+^{\mu}}{3}\left[1- \left(\frac{\mu}{\nu}\right) \frac{\Psi_+}{(1+\Delta s/s_+)^{4}} \left(\frac{\gamma(v_+)}{\gamma(c_b)}\right)^\nu \right],\\
    \P_{\rm driving} &= a_+ T_+^{\mu}\left[\frac{1}{3} +\left(\frac{\mu}{\nu}\right)\left(\alpha_+-\frac{1}{3}\right) \right].
\end{align}     
\end{subequations} 
The consequence of the entropy injection is to lower the effective value of the parameter $b$ by 
\bea 
b \to \frac{b}{(1+\Delta s/s_+)^4}, \qquad \Psi_+ \to \frac{\Psi_+}{(1+\Delta s/s_+)^4}. 
\eea 
This effectively increases the frictional pressure on the bubble wall, since $(1+\Delta s/s_+) > 1$. 

Assuming that $\P_{\rm dissipative}$, i.e., $\Delta \delta p$, is positive, one would confirm the idea that the LTE limit $\delta p \to 0, \Delta s \to 0$, gives a \emph{lower} bound on the pressure. Once $\delta p$ (or $\Delta\delta p$) is expressed in terms of $\Delta s/s_+$, then $\Delta s/s_+$ can be an additional parameter for hydrodynamic simulations. This is similar to that although $\alpha_n$ is computed from specific particle physics models, it is treated as a parameter in fluid dynamics~\cite{Jinno:2022mie}. 

\paragraph{$\Delta \delta p$ in the ultrarelativistic regime.}
One may get a feeling about $\Delta \delta p$ by considering the ultrarelativistic regime $\gamma_w\gg 1$. In this regime, we can use the collisionless/ballistic approximation. Then\footnote{Here, based on intuition, we take $f_i$ as the distribution function of the incoming particles right in front of the wall, which is $z$-independent. However, a rigorous analysis requires one to solve the Liouville equation.}
\begin{align}
\label{eq:deltaf-ultrarel}
\delta f_i = f_i - f_i^{\rm eq}(z;T(z)) &\approx f_i^{\rm in}(T_+)-f_i^{\rm eq}(z;T(z))\notag \\
&=\frac{1}{e^{\gamma_+ (|\vecp|+v_+ p^z)/T_+}\pm 1} -  \frac{1}{e^{\gamma(z)\left(\sqrt{\vecp^2 + m^2(z)}+v(z)p^z\right)/T(z)}\pm 1} > 0\,
\end{align}
when assuming $T(z)\approx T_+$ and $v(z)\approx v_+\approx 1$.
The signs $\pm$ correspond to the case of fermions/bosons entering the wall.
Above $f_i^{\rm in}$ are the distribution functions for the incoming particles. We thus obtain 
\bea 
    \Delta \delta p =\P_{\rm dissipative}= - \sum_i\int_{-\delta}^\delta \d z\, \underbrace{ \frac{\d m^2_i(\phi(z))}{\d z}  }_{<0}\underbrace{\left(\int \frac{\d^3{\bf p}}{(2\pi)^32E_i}\,\delta f_i\right)}_{> 0}  > 0.
\eea 
This shows that $\Delta \delta p$ adds more friction to the wall.

Outside of the ultrarelativistic regime, one might expect $f_i(z)\in [f_i^{\rm in}(T_+),f_i^{\rm eq}(z;T(z))]$ and $f_i^{\rm in}\approx f_i^{\rm in;eq}$. Assuming $T(z)\sim T_+$, $v(z)\sim v_+$, one again has $\delta f_i(z)\geq 0$ due to the suppression from the masses in $f_i^{\rm eq}(z;T(z))$ compared with $f_i^{\rm in; eq}(T_+)$. Thus, both $\Delta \delta p$ and $\Delta s$ increase the resistance from the plasma on the wall expansion. The limit $\Delta \delta p \to 0$ and $\Delta s \to 0$ is indeed a lower bound on the pressure.  

Further, in the ultrarelativistic regime we have $T_+=T_n$ and $f_i^{\rm in}=f^{\rm in;eq}_i$ (incoming particles are in thermal equilibrium at the nucleation temperature). Then one can recognize 
\begin{align}
\label{eq:93}
    \Delta \delta p= &- \sum_i\int_{-\delta}^\delta \d z\,  \frac{\d m^2_i(\phi(z))}{\d z}   \left(\int \frac{\d^3{\bf p}}{(2\pi)^32E_i}\,f^{\rm in;eq}_i(p;T_n)\right)\notag\\
    &+ \sum_i\int_{-\delta}^\delta \d z\,  \frac{\d m^2_i(\phi(z))}{\d z}   \left(\int \frac{\d^3{\bf p}}{(2\pi)^32E_i}\,f^{\rm eq}_i(p,z;T(z))\right)=\P_{\rm BM}-\P_{\rm LTE}.
\end{align}
This confirms that, for $\gamma_w\gg 1$,  $\P_{\rm friction}=\P_{\rm LTE}+\P_{\rm dissipative}\rightarrow \P_{\rm BM}$. Note that one in general does {\it not} have $\P_{\rm LTE}+\P_{\rm dissipative}=\P_{\rm BM}$ outside of the ultrarelativistic regime so that one cannot conclude that $\P_{\rm BM}$ is bigger than $\P_{\rm LTE}$ based on $\P_{\rm dissipative}>0$. To have $\P_{\rm friction}=\P_{\rm BM}$, we have assumed $(i)$ $f_i$ take the form of Eq.~\eqref{eq:deltaf-ultrarel}; $(ii)$ the particles in front of the wall are in thermal equilibrium; and $(iii)$ $T_+=T_n$. When any of these assumptions are not satisfied, one cannot conclude $\P_{\rm friction}=\P_{\rm BM}$. In particular, in the presence of the heating effects, $T_+>T_n$, and thus when the other two assumptions are satisfied, we have $\P_{\rm LTE}+\P_{\rm dissipative}>\P_{\rm BM}$.

\bibliographystyle{utphys}
\bibliography{ref}{}
\end{document}